\documentclass[traditabstract]{aa} 

\usepackage{graphicx}
\usepackage{txfonts}
\usepackage{natbib}
\newcommand{\kms}{{\hbox {km\thinspace s$^{-1}$}}}
\newcommand{\Lsun}{{\hbox {L$_\odot$}}}
\newcommand{\Msun}{{\hbox {M$_\odot$}}}

\newcommand{\cmt}{{\hbox {cm$^{-3}$}}}
\newcommand{\cmd}{{\hbox {cm$^{-2}$}}}
\newcommand{\hdo}{{\hbox {H$_{2}$O}}}

\def\t#1#2#3#4#5#6{{\hbox {$#1_{#2#3}-#4_{#5#6}$}}}

\def\13co{$^{13}$CO}
\def\c18o{C$^{18}$O}

\newcommand{\tgas}{{\hbox {$T_{\mathrm{gas}}$}}}

\newcommand{\tdust}{{\hbox {$T_{\mathrm{dust}}$}}}

\newcommand{\cwest}{{\hbox {$C_{\mathrm{west}}$}}}
\newcommand{\ceast}{{\hbox {$C_{\mathrm{east}}$}}}
\newcommand{\cext}{{\hbox {$C_{\mathrm{extended}}$}}}

\usepackage{color}
\begin{document}

   \title{Modeling the H$_2$O submillimeter emission in extragalactic
     sources}

   \author{E. Gonz\'alez-Alfonso \inst{1}, J. Fischer \inst{2},
     S. Aalto \inst{3}, N. Falstad \inst{3}
          }

   \institute{Universidad de Alcal\'a, Departamento de F\'{\i}sica
     y Matem\'aticas, Campus Universitario, E-28871 Alcal\'a de Henares,
     Madrid, Spain  
         \and
Naval Research Laboratory, Remote Sensing Division, 4555
     Overlook Ave SW, Washington, DC 20375, USA
         \and
Department of Earth and Space Sciences, Chalmers University of Technology,
Onsala Space Observatory, Onsala, Sweden 
}


 
   \authorrunning{Gonz\'alez-Alfonso}
   \titlerunning{Modeling the H$_2$O submillimeter emission in extragalactic
     sources}

  \abstract
   {Recent observational studies have shown that H$_2$O emission at (rest)
     submillimeter wavelengths is ubiquitous in infrared galaxies, both in the
     local and in the early Universe, suggestive of far-infrared pumping of
     H$_2$O by dust in warm regions. In this work, models are presented that
     show that $(i)$ the highest-lying H$_2$O lines ($E_{\mathrm{upper}}>400$ K)
     are formed in very warm ($T_{\mathrm{dust}}\gtrsim90$ K) regions and
     require high \hdo\ columns ($N_{\mathrm{H_2O}}\gtrsim3\times10^{17}$
     cm$^{-2}$), while lower lying lines can be efficiently excited with
     $T_{\mathrm{dust}}\sim45-75$ K and $N_{\mathrm{H_2O}}\sim(0.5-2)\times10^{17}$
     cm$^{-2}$; $(ii)$ significant collisional excitation of the lowest lying
     ($E_{\mathrm{upper}}<200$ K) levels, which enhances the overall
     $L_{\mathrm{H_2O}}$-$L_{\mathrm{IR}}$ ratios, is identified in sources
     where the ground-state para-\hdo\ \t111000\ line is detected in emission;
     $(iii)$ the \hdo-to-infrared ($8-1000$ $\mu$m) luminosity ratio is 
     expected to decrease with increasing \tdust\ for all lines with
     $E_{\mathrm{upper}}\lesssim300$ K, as has recently been reported in a
     sample of LIRGs, but increases with \tdust\ for the highest lying H$_2$O
     lines ($E_{\mathrm{upper}}>400$ K); $(iv)$ we find theoretical upper
     limits for $L_{\mathrm{H_2O}}/L_{\mathrm{IR}}$ in warm environments,
     owing to H$_2$O line saturation;  $(v)$ individual models are
     presented for two very different prototypical galaxies, the Seyfert 2
     galaxy NGC~1068 and the nearest ultraluminous infrared galaxy Arp~220,
     showing that the excited submillimeter H$_2$O emission is dominated by
     far-infrared pumping in both cases; $(vi)$ the
     $L_{\mathrm{H_2O}}$-$L_{\mathrm{IR}}$ correlation previously reported in
     observational studies indicates depletion or exhaustion time scales,
     $t_{\mathrm{dep}}=\Sigma_{\mathrm{gas}}/\Sigma_{\mathrm{SFR}}$, of
     $\lesssim12$ Myr for star-forming
     sources where lines up to $E_{\mathrm{upper}}=300$ K are detected, in
     agreement with the values previously found for (U)LIRGs from HCN 
     millimeter emission. 
We conclude that the submillimeter H$_2$O line emission other than the 
para-H$_2$O \t111000\ transition is pumped primarily by far-infrared
radiation, though some collisional pumping may contribute to the low-lying
para-H$_2$O \t202111\ line, and that collisional pumping of the 
para-$1_{11}$ and ortho-$2_{12}$ levels enhances the radiative
pumping of the higher lying levels.
}

   \keywords{Line: formation  
                 -- Galaxies: ISM  
                 -- Infrared: galaxies -- Submillimeter: galaxies
               }

   \maketitle
%

\section{Introduction}

With its high dipolar moment, extremely rich spectrum,
and high level spacing (in comparison to those of other molecules with
low-lying transitions at millimeter wavelengths), H$_2$O couples very well 
to the radiation field in warm 
regions that emit strongly in the far-IR. In extragalactic sources, excited
lines of \hdo\ at far-IR wavelengths ($\lambda<200$ $\mu$m) were
detected in absorption with the 
\emph{Infrared Space Telescope} (ISO) \citep{fis99,gon04,gon08}, and with
\emph{Herschel}/PACS \citep{pil10,pog10} in Mrk~231 \citep{fis10},
Arp~220 and NGC~4418 \citep[][G-A12]{gon12}. 
Modeling and analysis have demonstrated the 
ability of \hdo\ to be efficiently excited through absorption of far-IR
dust-emitted photons, thus providing a powerful method for studying the strength
of the far-IR field in compact/warm regions that are not spatially resolved at
far-IR wavelengths with current (or foreseen) technology. 

\emph{Herschel}/SPIRE \citep{gri10} has enabled the observation of
\hdo\ at submillimeter (hereafter submm, $\lambda>200$ $\mu$m)
wavelengths in local sources, where the excited (i.e., non-ground-state) lines
are invariably seen in emission. In Mrk~231, lines with $E_{\mathrm{upper}}$
up to $640$ K were detected \citep[][hereafter G-A10]{wer10,gon10}, with
strengths comparable to the CO lines. The \hdo\ lines have been also detected
in other local sources \citep{ran11,per13}, including the Seyfert 2 galaxy
NGC~1068 \citep[][S12]{spi12}. Furthermore, submm lines of \hdo\ have been
detected in a dozen of high-$z$ sources 
\citep{imp08,omo11,lis11,wer11,bra11,com12,lup12,both13}, even 
in a $z=6.34$ galaxy \citep{rie13}. Recently,  a striking correlation has
been found between the submm \hdo\ luminosity ($L_{\mathrm{H_2O}}$),
taken from the \t202111\ and \t211202\ lines, and the IR luminosity
($L_{\mathrm{IR}}$), including both local and high-$z$ ULIRGs
\citep[][hereafter O13]{omo13}. Using SPIRE spectroscopy of local
  IR-bright galaxies and published data from high-$z$ 
sources, the linear correlations between $L_{\mathrm{H_2O}}$ and
$L_{\mathrm{IR}}$ for five of the strongest lines, extending
over more than three orders of magnitude in IR luminosity, has recently been
confirmed \citep[][hereafter Y13]{yan13}. There
are hints of an increase in $L_{\mathrm{H_2O}}$ that is slightly faster than
linear with $L_{\mathrm{IR}}$ in some lines (\t211202\ and \t220211) and in
high-$z$ ULIRGs (O13). 
HCN is another key species that also shows a tight correlation with the
IR luminosity, even though the excitation of the $1-0$
transition is dominated by collisions with dense H$_2$ \citep{gao04a,gao04b}.

The increasing wealth of observations of \hdo\ at submm wavelengths in
both local and high-$z$ sources and the correlations discovered between
$L_{\mathrm{H_2O}}$ and $L_{\mathrm{IR}}$ require a more extended analysis in
parameter space than the one given in G-A10 for Mrk~231. In this work, models
are presented to constrain the physical and chemical conditions in the
submm \hdo\ emitting regions in warm
(U)LIRGs and to propose a general framework for
interpreting the \hdo\ submm emission in extragalactic sources.



\section{Excitation overview}
\label{sec:overview}

\begin{figure}
   \centering
   \includegraphics[width=8cm]{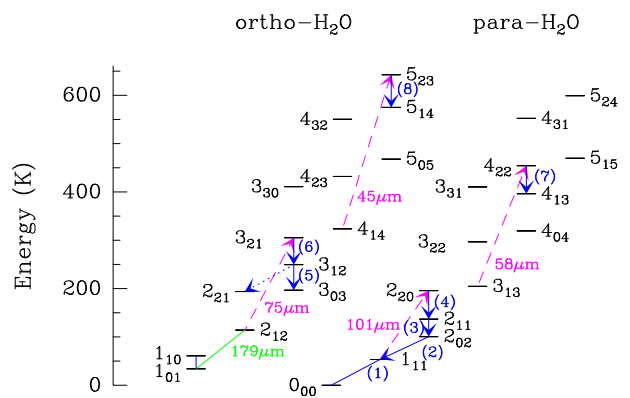}
   \caption{Energy level diagram of \hdo, showing the relevant
       \hdo\ lines at 
     submillimeter wavelengths with blue arrows, and the far-IR \hdo\ pumping
     (absorption) lines with dashed-magenta arrows. The lines are numbered
     as listed in Table~\ref{tab:tran}. The o-\hdo\ \t312221\ transition
     is not considered due to blending with CO (10-9) (G-A10), and the
       far-IR \t212101\ transition at $179.5$ $\mu$m discussed in the text is
       also indicated in green for completeness.
   }   
    \label{lev}
    \end{figure}

At submm wavelengths, H$_2$O responds to far-IR excitation by
emitting photons through a cascade proccess. This is illustrated in
Fig.~\ref{lev}, where four far-IR pumping lines (at 101, 75, 58, and 45 $\mu$m)
account for the radiative excitation of the submm lines (G-A10). The line
parameters are listed in Table~\ref{tab:tran}, where we use the numerals
  $1-8$ to denote the submm lines. Lines $2-4$, $5-6$, $7$, and $8$ are pumped
through the $101$\footnote{This line lies within the PACS 100 $\mu$m
  gap, but was detected in Arp~220 and Mrk~231 with {\it ISO}
  \citep{gon04,gon08}.}, $75$, $58$, and $45$ $\mu$m far-IR transitions,
respectively.

The ground-state line 1 has no analog pumping mechanism, so that the upper
$1_{11}$ level can only be excited through absorption of a photon in the same
transition (at $269$ $\mu$m) or through a collisional event. In the absence of
significant collisional excitation, and if approximate spherical symmetry
holds, line 1 will give negligible absorption or emission above the continuum 
(regardless of line opacity) if the continuum opacity at $269$ $\mu$m is
low or will be detected in absorption for significant $269$ $\mu$m continuum
opacities\footnote{This is 
  analogous to the behavior of the OH 119 $\mu$m doublet, see \cite{gon14}.}.
This is supported by the SPIRE spectrum of Arp~220, in which line 1 is
observed in absorption \citep{ran11} and high submm continuum
opacities are inferred \citep{gon04,dow07,sak08}. Collisional excitation and
thus high densities and gas temperatures are then expected in sources where
line 1 is detected in emission (10 sources among 176, Y13), as in NGC~1068
(S12; see also App.~\ref{appa}). 
Line 1 can then be collisionally excited in regions where the other lines
do not emit owing to  weak far-IR continuum;
this effect has recently been observed in the intergalactic filament in the
Stephan's Quintet \citep{app13}. 

If collisional excitation of the $1_{11}$ and $2_{12}$ levels dominates over
absorption of dust photons at $269$ and $179$ $\mu$m (i.e., in very optically
thin and/or high density sources), the submm \hdo\ lines $2-6$ will be
boosted because these $1_{11}$ and $2_{12}$ levels are the {\it base levels}
from which the 101 and 75 $\mu$m radiative pumping cycles operate
(Fig.~\ref{lev}). In addition, in regions of low continuum opacities but
  warm gas, collisional excitation of the para-\hdo\ level $2_{02}$ from the
  ground $0_{00}$ state can significantly enhance the emission of line 2.
Therefore, the \hdo\ submm emission
depends in general on both the far-IR radiation density in the emitting
region and the possible collisional excitation of the low-lying
levels ($1_{11}$, $2_{12}$, and $2_{02}$). Lines $7-8$ require 
strong far-IR radiation density not only at $58-45$ $\mu$m, but also at longer
wavelengths, together with high \hdo\ column densities ($N_{\mathrm{H_2O}}$) in
order to significantly populate the lower backbone $3_{13}$ and $4_{14}$ levels.

   \begin{table}
      \caption[]{H$_2$O transitions at $\lambda>200$ $\mu$m considered in this
        work.} 
         \label{tab:tran}
          \begin{tabular}{clccc}   
            \hline
            \noalign{\smallskip}
N & Transition  & $E_{\mathrm{upper}}$ & $\lambda_{\mathrm{rest}}$ & $A_{ul}$   \\  
& &  (K)  & ($\mu$m) & (s$^{-1}$) \\
            \noalign{\smallskip}
            \hline
            \noalign{\smallskip}
1 & \hdo\ \, \t111000\ & $53$  & $269.27$ & $0.018$  \\
2 & \hdo\ \, \t202111\ & $101$ & $303.46$ & $0.006$  \\
3 & \hdo\ \, \t211202\ & $137$ & $398.64$ & $0.007$  \\
4 & \hdo\ \, \t220211\ & $196$ & $243.97$ & $0.019$  \\
5 & \hdo\ \, \t312303\ & $249$ & $273.19$ & $0.016$  \\
6 & \hdo\ \, \t321312\ & $305$ & $257.79$ & $0.023$  \\
7 & \hdo\ \, \t422413\ & $454$ & $248.25$ & $0.028$  \\
8 & \hdo\ \, \t523514\ & $642$ & $212.53$ & $0.043$  \\
            \noalign{\smallskip}
            \hline
         \end{tabular} 
   \end{table}

   \begin{table}
      \caption[]{Model parameters.} 
         \label{tab:par}
          \begin{tabular}{lcc}   
            \hline
            \noalign{\smallskip}
Parameter & Explored range  & Best fit to  \\
          &                 & HII+mild AGN$^{\mathrm{a}}$ \\  
            \noalign{\smallskip}
            \hline
            \noalign{\smallskip}
\tdust\ (K)                     &   $35-115$   & $45-75$   \\
$\tau_{100}$                    &    $0.02-12$   & $0.05-0.2$   \\
$N_{\mathrm{H_2O}}/\Delta V$ (cm$^{-2}$/(km s$^{-1}$))   &
$3\times10^{14}-10^{16}$ & $(0.5-2)\times10^{15}$ \\ 
$\Delta V^{\mathrm{c}}$ (km s$^{-1}$) &    $100$  &   $^{\mathrm{b}}$ \\
$n_{\mathrm{H_2}}$ (cm$^{-3}$)    &    $2\times10^{4}-3\times10^{5}$ &
$\lesssim3\times10^5$      \\
\tgas\ (K)                      &    $100-200$ &  $^{\mathrm{b}}$   \\
            \noalign{\smallskip}
            \hline
         \end{tabular} 
\begin{list}{}{}
\item[$^{\mathrm{a}}$] Best fit values for HII+mild AGN sources 
  (optically classified star-formation dominated galaxies with possible
    mild AGN contribution, see Y13 and Sect.~\ref{sec:summary}) for which lines
  $2-6$ are detected, but lines $7-8$ are undetected.
\item[$^{\mathrm{b}}$] Parameter not well constrained. 
\item[$^{\mathrm{c}}$] FWHM velocity dispersion of the dominant \hdo\ emitting
  structures, in our models equal to $1.67 V_{\mathrm{turb}}$.
\end{list}
   \end{table}

\section{Description of the models}

The basic models for \hdo\ were described in G-A10 (see also
references therein). Summarizing, we assume a simple spherically symmetric
source with uniform physical properties (\tdust, \tgas, gas and dust
densities, \hdo\ abundance), where gas and dust are assumed to be
mixed. We only consider the far-IR radiation field generated within the
modeled source, ignoring the effect of external fields. The source 
is divided into a set of spherical shells where the 
statistical equilibrium level populations are calculated. The models
are non-local, including line and continuum opacity effects. We assume
an \hdo\ ortho-to-para ratio of 3. Line
broadening is simulated by including a microturbulent velocity
($V_{\mathrm{turb}}$), for which the FWHM velocity dispersion is
$\Delta V=1.67V_{\mathrm{turb}}$. No systemic motions are included.

\subsection{Mass absorption coefficient of dust}
\label{sec:kabs}

\begin{figure}
   \centering
   \includegraphics[width=8cm]{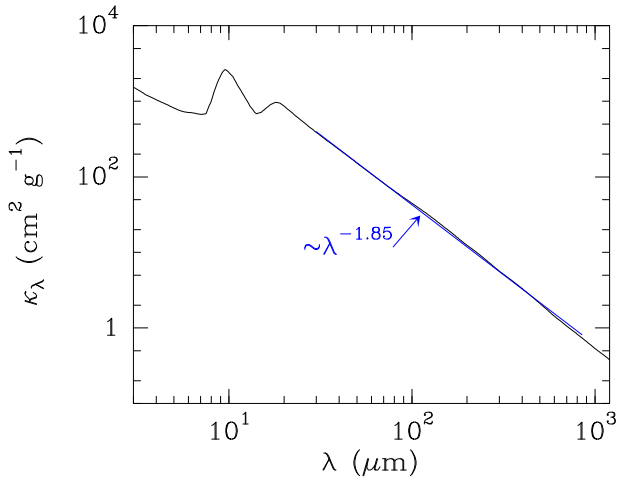}
   \caption{Adopted mass absorption coefficient of dust as a function of
     wavelength. The dust emission is simulated by using a mixture
    of silicate and amorphous carbon grains with optical constants from
    \cite{dra85} and \cite{pre93}. As shown by the fitted blue line, the
    emissivity index from the far-IR to millimeter wavelengths is
    $\beta=1.85$. 
   }   
    \label{kabs}
    \end{figure}

The black curve in Fig.~\ref{kabs} shows the dust mass opacity
coefficient used in the current and our past models
\citep{gon08,gon10,gon12,gon13,gon14}. Our values at 125
and 850 $\mu$m are $\kappa_{125}=30$ cm$^2$ g$^{-1}$ and $\kappa_{850}=0.7$ cm$^2$
g$^{-1}$, in good agreement with those derived by \cite{dun03}. Adopting a
gas-to-dust ratio of $X=100$ by mass, and using $\kappa_{100}=44.5$ cm$^2$
g$^{-1}$, the column density of H nuclei is
\begin{equation}
N_{\mathrm{H}}=\frac{X\,\tau_{100}}{m_{\mathrm{H}}\,\kappa_{100}}=
1.3\times10^{24}\,\tau_{100} \, \mathrm{cm^{-2}},
\label{eq:nh}
\end{equation}
where $\tau_{100}$ is the continuum optical depth at 100 $\mu$m. 

For this adopted dust composition, the fit across the far-IR to submm (blue
line in Fig.~\ref{kabs}) indicates an emissivity index of
$\beta=1.85$, slightly steeper than the $\beta=1.5-1.6$ values favored by
\cite{kov10} and \cite{cas12}. The \hdo\ excitation is sensitive to the
dust emission over a range of wavelengths (from $45$ to $270$ $\mu$m), 
but we find that our results on $L_{\mathrm{H_2O}}/L_{\mathrm{IR}}$ are
insensitive to $\beta$ for $\beta$ values above $1.5$
(Sect.~\ref{sec:lh2olir-tdust}).

\subsection{Model parameters}
\label{sec:param}

As listed in Table~\ref{tab:par}, the model parameters we have chosen to characterize the physical conditions in the emitting regions are
  \tdust, the continuum optical depth at 100 $\mu$m along a radial path
($\tau_{100}$), the corresponding \hdo\ column density per unit of
  velocity interval ($N_{\mathrm{H_2O}}/\Delta V$), the velocity
  dispersion $\Delta V$, \tgas, and the H$_2$ density
($n_{\mathrm{H_2}}$). Fiducial numbers for some of these parameters are 
$\tau_{100}=0.1$, $\Delta V=100$ \kms, $T_{\mathrm{gas}}=150$ K, and 
$n_{\mathrm{H_2}}=3\times10^5$ \cmt. Collisional rates with H$_2$ were taken
from \cite{dub09} and \cite{dan11}. Our relevant results are the
line-flux ratios ($F_i/F_j$) and the luminosity ratios\footnote{We
denote $L_{\mathrm{H_2O}}$ as the \hdo\ luminosity {\it of a given generic}
\hdo\ submm line, while $L_{i}$ is the luminosity of the \hdo\ line $i$
(numbering in Table~\ref{tab:tran}). The \hdo\ line fluxes $F_i$ are given
in Jy km s$^{-1}$. $L_{\mathrm{IR}}$ is the $8-1000$ $\mu$m luminosity.}
$L_{\mathrm{H_2O}}/L_{\mathrm{IR}}$.
We also explore models where collisions are ignored, appropriate for
low-density regions ($n_{\mathrm{H_2}}\lesssim 10^4$ \cmt), for which $F_i/F_j$
only depend on \tdust, $\tau_{100}$, and $N_{\mathrm{H_2O}}/\Delta V$, while
$L_{\mathrm{H_2O}}/L_{\mathrm{IR}}$ depends in addition on $\Delta V$.

Depending on the values of the above parameters, our models can be
interpreted in terms of a single source or are better applied to each of an
ensemble of clouds within a clumpy distribution. The radius of the modeled
source is 
\begin{equation}
R= N_{\mathrm{H}}/n_{\mathrm{H}} = 0.21 \times
\left( \frac{\tau_{100}}{0.1} \right) \times
\left( \frac{10^5 \, \mathrm{cm^{-3}}}{n_{\mathrm{H_2}}} \right) \, \mathrm{pc},
\label{eq:r}
\end{equation}
where Eq.~(\ref{eq:nh}) has been applied. The corresponding IR luminosity can
be written as $L_{\mathrm{IR}}=4\pi R^2 \sigma T_{\mathrm{dust}}^4\, \gamma$,
where $\gamma(T_{\mathrm{dust}},\tau_{100})\le1$ accounts for the departure from a
blackbody emission due to finite optical depths, ranging from $\gamma=0.2$
for $T_{\mathrm{dust}}=50$ K and $\tau_{100}=0.1$ to $\gamma=0.9$ for
$T_{\mathrm{dust}}=95$ K and $\tau_{100}=1$. In physical units,
\begin{equation}
 L_{\mathrm{IR}} = 1.4\times10^5 \times
\left( \frac{\tau_{100}}{0.1} \right)^2 \times
\left( \frac{10^5 \, \mathrm{cm^{-3}}}{n_{\mathrm{H_2}}} \right)^2 \times
\left( \frac{T_{\mathrm{dust}}}{55 \, \mathrm{K}} \right)^4 \times
\left( \frac{\gamma}{0.2} \right)
\label{eq:lir}
\end{equation}
in \Lsun, indicating that a model with $\tau_{100}\sim0.1$ and moderate
$T_{\mathrm{dust}}$ should be considered as one of an ensemble of clumps to
account for the typically observed IR luminosities of $\gtrsim10^{11}$ \Lsun\
(Y13). For very warm ($T_{\mathrm{dust}}\sim90$ K) and optically thick
($\tau_{100}\sim1$) sources with low average densities 
($n_{\mathrm{H_2}}=3\times10^3$ \cmt), Eq.~(\ref{eq:lir}) gives 
$L_{\mathrm{IR}} \sim 5\times10^{11}$ \Lsun\ and the model can be applied to
a significant fraction of the circumnuclear region of galaxies where the
clouds may have partially lost their individuality \citep{dow98}.

The velocity dispersion $\Delta V$ in our models can be related to the
  velocity gradient used in escape probability methods as
$dV/dr\sim \Delta V/(2R)$, and using Eq.~(\ref{eq:r})
\begin{equation}
dV/dr \sim 238 \times
\left( \frac{\Delta V}{100 \, \mathrm{km \, s^{-1}}} \right) \times
\left( \frac{0.1}{\tau_{100}} \right) \times
\left( \frac{n_{\mathrm{H_2}}}{10^5 \, \mathrm{cm^{-3}}} \right)
\, \mathrm{km \, s^{-1} \, pc^{-1}.}
\label{eq:dvdr}
\end{equation}
Defining $K_{\mathrm{vir}}$ as the ratio of the velocity gradient relative to
that expected in gravitational virial equilibrium,
$K_{\mathrm{vir}}=(dV/dr)/(dV/dr)_{\mathrm{vir}}$, and using
$(dV/dr)_{\mathrm{vir}}\sim 10 \times (n_{\mathrm{H_2}}/10^5 \, 
\mathrm{cm^{-3}})^{1/2}$ $\mathrm{km \, s^{-1} \, pc^{-1}}$
\citep{bry96,gol01,pap07,hay12}, we obtain
\begin{equation}
K_{\mathrm{vir}} \sim 23.8  \times
\left( \frac{\Delta V}{100 \, \mathrm{km \, s^{-1}}} \right) \times
\left( \frac{0.1}{\tau_{100}} \right) \times
\left( \frac{n_{\mathrm{H_2}}}{10^5 \, \mathrm{cm^{-3}}} \right)^{1/2}.
\label{eq:kvir}
\end{equation}
Values of $K_{\mathrm{vir}}$ significantly above 1 and up to $\sim20$, 
indicating non-virialized phases, have been inferred in luminous IR
galaxies from both low- and high-$J$ CO lines
\citep[e.g.,][]{pap99,pap07,hay12}. For clarity, the velocity dispersion is 
rewritten in terms of $K_{\mathrm{vir}}$ as
\begin{equation}
\Delta V = 42 \times 
\left( \frac{K_{\mathrm{vir}}}{10} \right) \times
\left( \frac{\tau_{100}}{0.1} \right) \times
\left( \frac{10^5 \, \mathrm{cm^{-3}}}{n_{\mathrm{H_2}}} \right)^{1/2} \,
\mathrm{km \, s^{-1}},
\label{eq:deltavvir}
\end{equation}
which shows that, for compact and dense clumps ($\tau_{100}=0.1$,
$n_{\mathrm{H_2}}=3\times10^5$ \cmt), $\Delta
V\sim25\times(K_{\mathrm{vir}}/10)$ \kms\ and the typical observed linewidths
($\sim300$ \kms) are caused by the galaxy rotation pattern and velocity
dispersion of clumps. In contrast, for optically thick sources with low
densities ($\tau_{100}=1$, $n_{\mathrm{H_2}}\lesssim10^4$ \cmt), $\Delta
V\gtrsim130$ km s$^{-1}$ is required for $K_{\mathrm{vir}}\gtrsim1$.

Instead of calculating $\Delta V$ for each model according to
  eq.~(\ref{eq:deltavvir}), which would involve a ``universal''
  $K_{\mathrm{vir}}$ independent of the source characteristics\footnote{
We may expect $K_{\mathrm{vir}}>1$ for clouds in a clumpy distribution due to
the gravitational potential of the galaxy and external pressure \citep{pap99},
but $K_{\mathrm{vir}}\sim1$ may be more appropriate for sources where the
clouds have coalesced and the resulting (modeled) structure can be considered
more isolated. However, $K_{\mathrm{vir}}>1$ in case of prominent outflows.},
we have used $\Delta V=100$ \kms\ for comparison purposes between
models (in Sect.~\ref{sec:corr} we also consider models with constant
$K_{\mathrm{vir}}$). Nevertheless, results can be easily rescaled to any
other value of $\Delta V$ as follows.  
For given \tdust\ and $\tau_{100}$, the relative level populations,
the line opacities, and thus the \hdo\ line-flux ratios ($F_i/F_j$) depend on  
$N_{\mathrm{H_2O}}/\Delta V$, while the luminosity ratios
$L_{\mathrm{H_2O}}/L_{\mathrm{IR}}$ are proportional to $\Delta V$. Therefore,
for any $\Delta V$, identical results for $F_i/F_j$ are obtained with the
substitution 
\begin{equation}
N_{\mathrm{H_2O}} \rightarrow  N_{\mathrm{H_2O}} \times  
\left( \frac{\Delta V}{100 \, \mathrm{km \, s^{-1}}} \right),
\end{equation}
while $L_{\mathrm{H_2O}}/L_{\mathrm{IR}}$ should be scaled as 
\begin{equation}
\frac{L_{\mathrm{H_2O}}}{L_{\mathrm{IR}}} \rightarrow 
\frac{L_{\mathrm{H_2O}}}{L_{\mathrm{IR}}} \times 
\left( \frac{\Delta V}{100 \, \mathrm{km \, s^{-1}}} \right).
\end{equation}

   \begin{figure*}
   \centering
   \includegraphics[width=15.0cm]{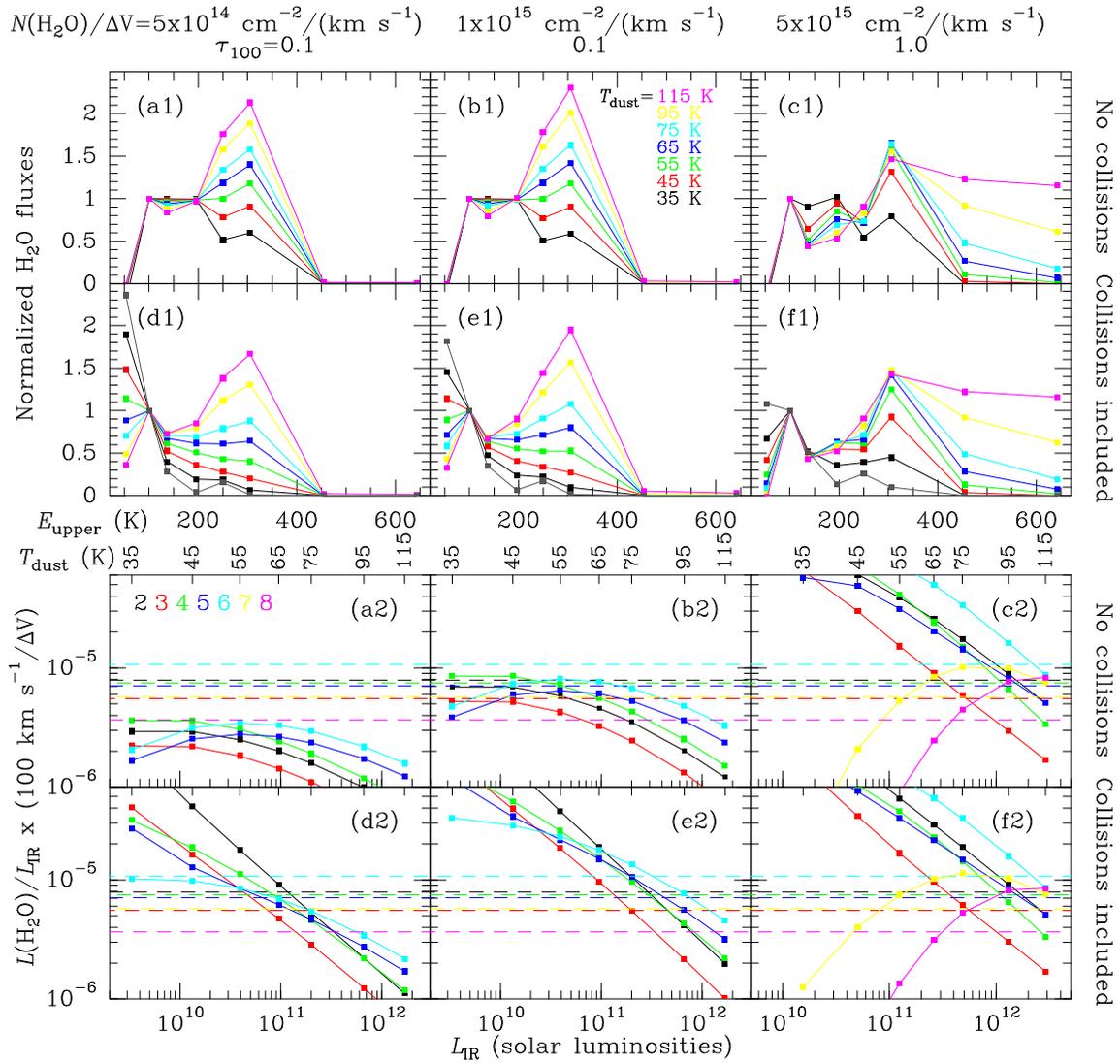}
   \caption{Relevant model results for the normalized \hdo\ SLED {\bf (a1-f1)},
     and for the $L_{\mathrm{H_2O}}/L_{\mathrm{IR}}$ ratios (for $\Delta
     V=100$ \kms) as a function of \tdust\ and $L_{\mathrm{IR}}$ (assuming a
     source of $R_{\mathrm{eff}}=100$ pc, {\bf a2-f2}). 
     In panels a1-f1, model results for lines $1$ to $8$
       (Table~\ref{tab:tran}) are shown from left to right. 
     Values for $N_{\mathrm{H_2O}}/\Delta V$ and $\tau_{100}$ are 
     indicated at the top of the figure. The different colors in panels {\bf
       a1-f1} indicate different \tdust, as labeled in {\bf b1}, while they
     indicate different lines in panels {\bf a2-f2} (labeled in {\bf a2}, see
     Table~\ref{tab:tran}). Models with
     collisional excitation ignored ({\bf a-c}), and with collisions included
     for $n_{\mathrm{H_2}}=3\times10^5$ \cmt\ and $T_{\mathrm{gas}}=150$ K
     ({\bf d-f}) are shown. The gray lines/symbols in panels {\bf d1-f1} show
     model results that ignore radiative pumping (i.e., only collisional
     excitation). Collisional excitation has the overall effect of
     enhancing the low-lying lines (1 and 2) relative to the others and of
     increasing the $L_{\mathrm{H_2O}}/L_{\mathrm{IR}}$ ratios of all lines
     (see text). The dashed lines in panels {\bf a2-f2} indicate the average
     $L_{\mathrm{H_2O}}/L_{\mathrm{IR}}$ ratios reported by Y13. When
     compared with observations, the modeled $L_{\mathrm{IR}}$ values should
     be considered a fraction of the observed IR luminosities, because single
     temperature dust models are unable to reproduce the observed SEDs 
     (Sect.~\ref{sed}); the \hdo\ submm emission traces warm regions of
     luminous IR galaxies (see text). 
   }   
    \label{mod}
    \end{figure*}

Both the line-flux ratios ($F_i/F_j$) and the luminosity ratios
$L_{\mathrm{H_2O}}/L_{\mathrm{IR}}$ are independent of the number of
  clumps ($N_{\mathrm{cl}}$) if the model parameters (\tdust, $\tau_{100}$,
  \tgas, $n_{\mathrm{H_2}}$, $N_{\mathrm{H_2O}}/\Delta V$, and $\Delta V$)
  remain the same for the cloud average. With the effective source radius
  defined as $R_{\mathrm{eff}}=N_{\mathrm{cl}}^{1/2}R$, both the line and continuum
  luminosities scale as $\propto R_{\mathrm{eff}}^2$. Therefore, if the
effective source size is changed and all other
parameters are kept constant, a linear correlation between each
$L_{\mathrm{H_2O}}$ and $L_{\mathrm{IR}}$ is naturally generated, regardless
of the excitation mechanism of  \hdo. (For reference, however, all absolute
luminosities below are given for $R_{\mathrm{eff}}=100$ pc.) The question,
then, is what range of dust and gas parameters characterizes the sources for
which the observed nearly linear correlations in lines $2-6$ 
(O13, Y13) are observed. The detection rates of lines $1$, $7$, and $8$ are
relatively low, but the same trend is observed in the few sources where they
are detected (Y13). 

In the following sections, the general results of our models are presented,
while specific fits to two extreme sources, Arp~220 and NGC~1068, 
are discussed in Appendix~\ref{appa}.

\section{Model results}
\label{sec:results}

\subsection{General results}
\label{sec:general}

\begin{figure}
   \centering
   \includegraphics[width=9cm]{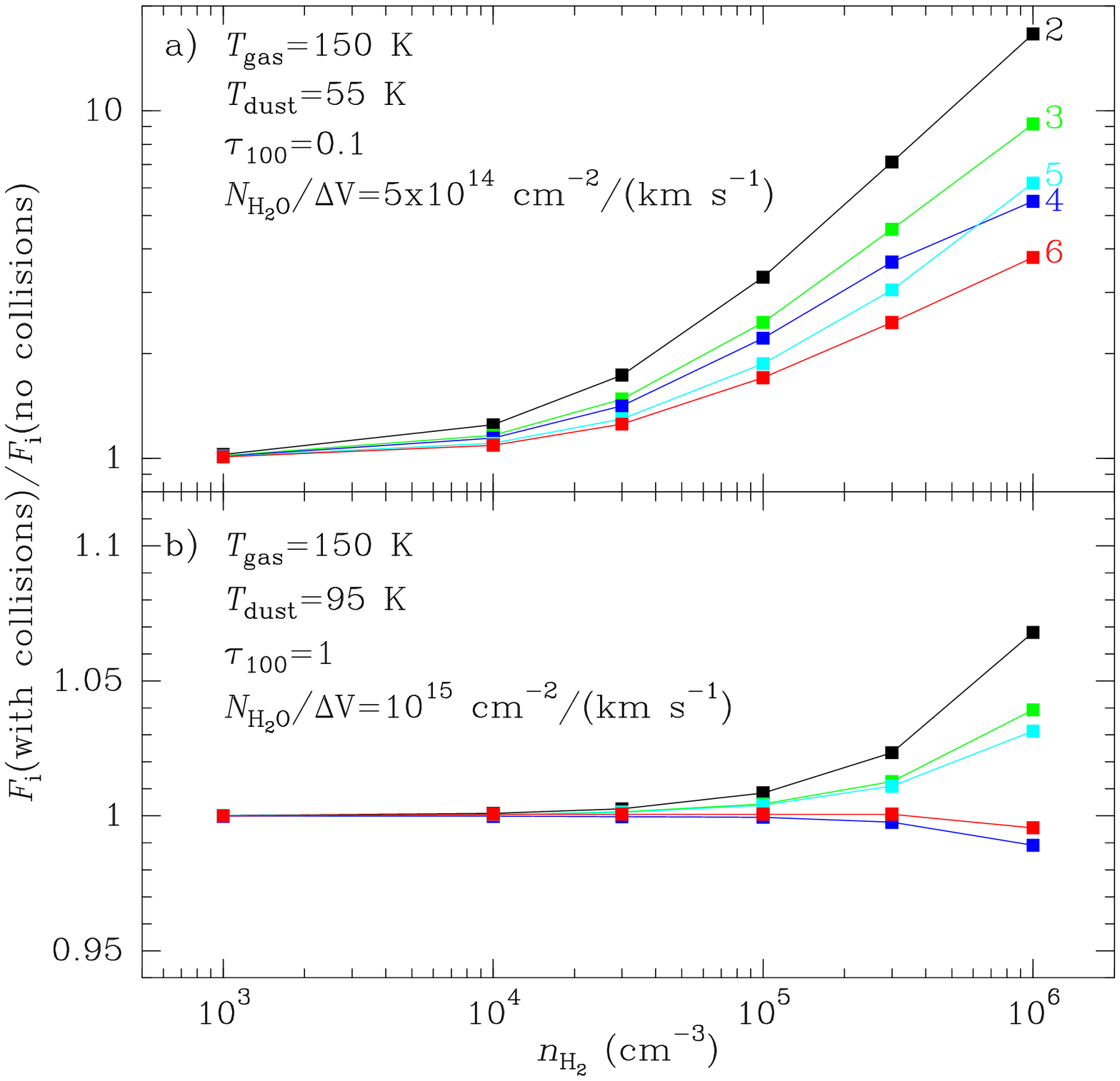}
   \caption{Effect of collisional excitation on the \hdo\ fluxes of the submm
     lines $2-6$ as a function of $n_{\mathrm{H_2}}$. The ordinates show the
     calculated line fluxes relative to the model that ignores collisional
     excitation. $T_{\mathrm{gas}}=150$ K is adopted in all models.
     {\bf a)} In the case of moderate \tdust\ and low $\tau_{100}$, collisional
     excitation has a strong impact on the \hdo\ fluxes at $n_{\mathrm{H_2}}$
     of $\mathrm{a\, few\times10^4}$ \cmt, especially on line 2. {\bf b)}
     Collisional excitation is negligible for high $\tau_{100}$ and very warm
     \tdust\ (note the difference in ordinate scales in {\bf a)} and {\bf b)}).
   }   
    \label{colis}
    \end{figure}

In Fig.~\ref{mod}, model results are shown in which \tdust\ is varied
from 35 to 115 K, $N_{\mathrm{H_2O}}/\Delta V$ from $5\times10^{14}$ to
$5\times10^{15}$ $\mathrm{cm^{-2}/(km\,s^{-1})}$, $\tau_{100}$ from $0.1$ to
$1.0$, and where collisional excitation with $n_{\mathrm{H_2}}=3\times10^5$
\cmt\ and $T_{\mathrm{gas}}=150$ K is excluded (a-c) or included (d-f). Panels
a1-f1 (top) show the expected SLED normalized to line 2, and panels a2-f2 
(bottom) plot the corresponding 
$L_{\mathrm{H_2O}}/L_{\mathrm{IR}}\times (100 \, \mathrm{km\,s^{-1}}/\Delta
V)$ ratios as a function of \tdust\ and $L_{\mathrm{IR}}$ (for
$R_{\mathrm{eff}}=100$ pc; all points would move horizontally for 
different $R_{\mathrm{eff}}$). The effect of collisional excitation is also
illustrated in Fig.~\ref{colis}, where the \hdo\ submm fluxes of lines $2-6$
relative to those obtained ignoring collisional excitation are
plotted as a function of $n_{\mathrm{H_2}}$ for $T_{\mathrm{gas}}=150$ K.

The first conclusion that we infer from Fig.~\ref{mod}a1-f1 is that the
relative fluxes of lines $5-6$ generally increase with increasing \tdust. 
These lines are pumped through the \hdo\ transition at
$\lambda\sim75$ $\mu$m (Fig.~\ref{lev}), thus requiring warmer dust than
lines $2-4$, which are pumped through absorption of 100 $\mu$m photons.
The SLEDs obtained with $T_{\mathrm{dust}}<45$ K yield $F_4$
significantly above $F_6$, and are thus unlike those observed 
in most (U)LIRGs (Y13).
The two peaks in the \hdo\ SLED (in lines 2 and 6) generally found in (U)LIRGs
(Y13) indicate that the submm \hdo\ emission essentially samples
regions with $T_{\mathrm{dust}}\gtrsim45$ K. Significant collisional excitation
enhances line 4 relative to line 6 (Fig.~\ref{mod}d1-f1), thus
aggravating the discrepancy between the $T_{\mathrm{dust}}<45$ K models and
the observations.

Lines $7-8$ provide stringent constraints on \tdust, $\tau_{100}$,
and $N_{\mathrm{H_2O}}/\Delta V$. Since line 6 is still easily excited even 
with moderately warm $T_{\mathrm{dust}}\sim55$ K, the 8/6 and 7/6 ratios are 
good indicators of whether very warm dust ($>80$ K) is exciting \hdo. 
Sources where lines $7-8$ are detected (e.g., Mrk~231, Arp~220, 
and APM~08279) can be considered ``very warm'' on these grounds, with
$N_{\mathrm{H_2O}}/\Delta V\gtrsim3\times10^{15}$
$\mathrm{cm^{-2}/(km\,s^{-1})}$. Sources where lines 
$7-8$ are not detected to a significant level, but where the SLED still shows a
second peak in line 6, are considered ``warm'', i.e. with \tdust\ varying
between $\sim45$ and 80 K, and $N_{\mathrm{H_2O}}/\Delta
V\sim(5-20)\times10^{14}$ $\mathrm{cm^{-2}/(km\,s^{-1})}$.    

Sources in which lines $2-6$ are not detected to a significant level,
that do not show a second peak in line 6, or for which the
\hdo\ luminosities are well below the observed
$L_{\mathrm{H_2O}}-L_{\mathrm{IR}}$ correlation are considered 
``cold''. These sources are characterized by 
very optically thin and extended continuum emission, and/or with low
$N_{\mathrm{H_2O}}$ (these properties likely go together). Such sources
include starbursts like M82 (Y13), where the continuum is generated in PDRs
and are physically very different from the properties of ``very warm'' sources
like Mrk~231 (G-A10).  

In the models that neglect collisional excitation (a1-c1), line 1 is 
predicted to be in absorption, transitioning to emission in
warm/dense regions where it is collisionally excited (d1-f1), as previously
argued. Its strength will also depend on the continuum opacity, which should
be low enough to allow the line to emit above the continuum. 
Direct collisional excitation from the ground state in regions with warm
  gas but low $\tau_{100}$ efficiently populates level $2_{02}$, so that the
2/3, 2/4, 2/5, and 2/6 ratios strongly increase with increasing
$n_{\mathrm{H_2}}$ (Fig.~\ref{colis}a).  
As advanced in Sect.~\ref{sec:overview}, collisional
excitation also boosts all other submm lines for moderate \tdust\ owing to
efficient pumping of the base levels $2_{12}$ and $1_{11}$; radiative
trapping of photons emitted in the ground-state transitions increases
the chance of absorption of continuum photons in the
$101$ and $75$ $\mu$m transitions. Nevertheless, collisional excitation is
negligible for high $\tau_{100}$ and high \tdust\ (Fig.~\ref{colis}b).

\subsection{Predicted line ratios}
\label{lratios}

\begin{figure}
   \centering
   \includegraphics[width=8.3cm]{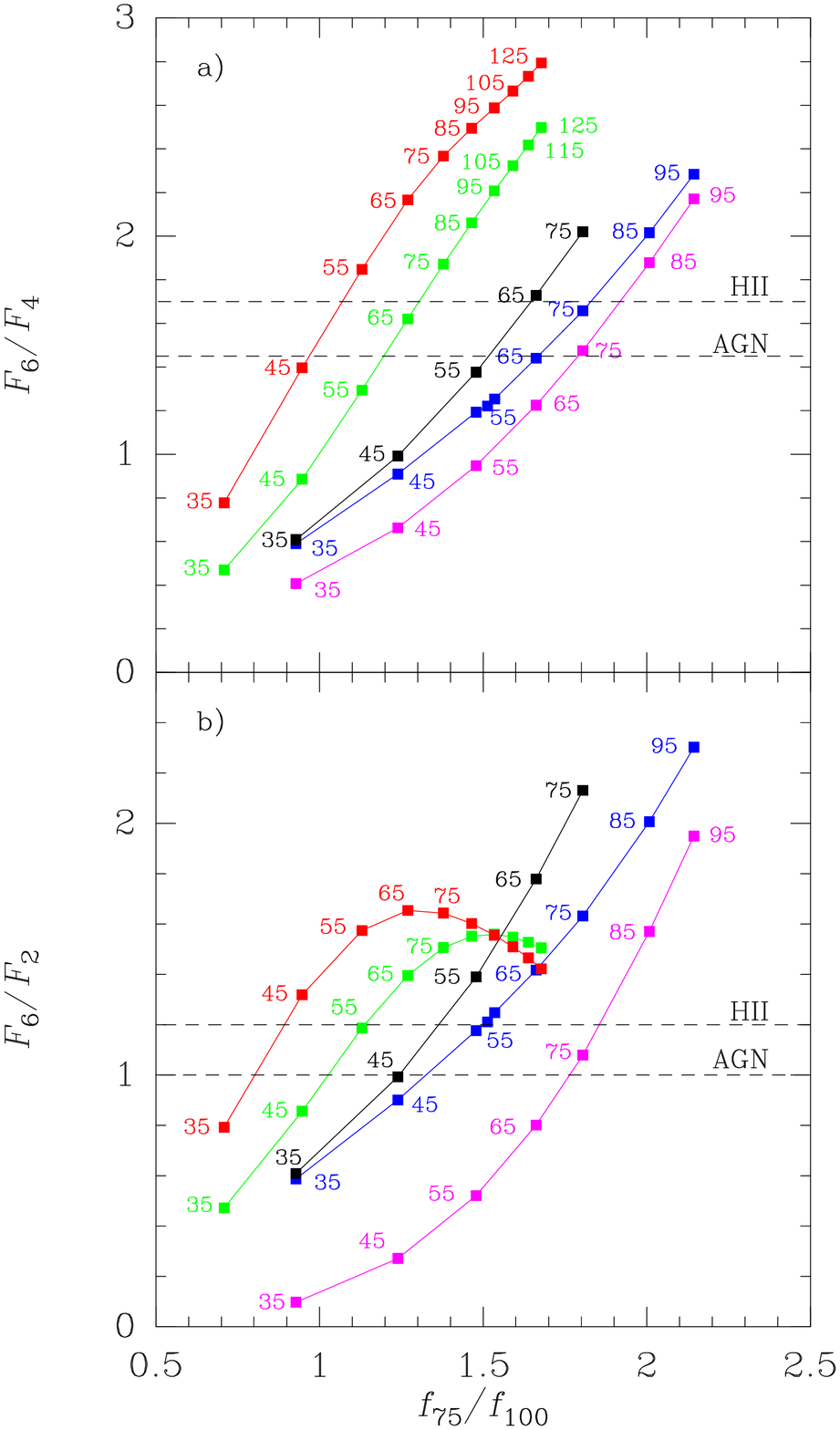}
   \caption{{\bf a)} $F_6/F_4$ (o\hdo\ \t321312-to-p\hdo\ \t220211) and 
     {\bf b)} $F_6/F_2$ (o\hdo\ \t321312-to-p\hdo\ \t202111) line flux
     ratios as a function of the 75-to-100 $\mu$m far-IR color. Blue symbols:
     $N_{\mathrm{H_2O}}/\Delta V=10^{15}$ $\mathrm{cm^{-2}/(km\,s^{-1})}$ and
     $\tau_{100}\leq0.1$; black: 
     $N_{\mathrm{H_2O}}/\Delta V=5\times10^{15}$
     $\mathrm{cm^{-2}/(km\,s^{-1})}$ and $\tau_{100}\leq0.1$; red: 
     $N_{\mathrm{H_2O}}/\Delta V=5\times10^{15}$
     $\mathrm{cm^{-2}/(km\,s^{-1})}$ and $\tau_{100}=1.0$; green: 
     $N_{\mathrm{H_2O}}/\Delta V=10^{15}$ $\mathrm{cm^{-2}/(km\,s^{-1})}$ and
     $\tau_{100}=1.0$; magenta: same as 
     blue symbols but with collisional excitation included with
     $T_{\mathrm{gas}}=150$ K and $n_{\mathrm{H_2}}=3\times10^{5}$ \cmt. The
     small numbers close to the symbols indicate the value of \tdust. The
     observed averaged ratios for strong-AGN and HII+mild-AGN sources (Y13)
     are indicated with dashed horizontal lines, and indicate that the
     regions probed by the \hdo\ submm emission are characterized by
     warm dust ($T_{\mathrm{dust}}\gtrsim45$ K).  
   }   
    \label{ratio62}
    \end{figure}

In sources where lines 7 and 8 are not detected,
the 6/4 flux ratio is the most direct indication of the hardness of the
far-IR radiation field seen by the \hdo\ gas responsible for the observed
emission. Since line 4 is pumped through absorption of 101 $\mu$m
photons and line 6 by 75 $\mu$m photons (Fig.~\ref{lev}), one may expect a
correlation between the 6/4 ratio and the 75-to-100 $\mu$m far-IR color,
$f_{75}/f_{100}$. As shown in Fig.~\ref{ratio62}a, our models indeed show a
steep increase in the 6/4 ratio with \tdust\ for fixed
$\tau_{100}$ and $N_{\mathrm{H_2O}}$. The averaged observed 6/4 ratio of 
$\approx1.45-1.7$ in strong-AGN and HII+mild-AGN sources (Y13) indicates,
assuming an optically thin continuum (Fig.~\ref{ratio62}a),
$T_{\mathrm{dust}}\approx55-75$ K and 
$f_{75}/f_{100}=1.5-1.8$. For the case of high $\tau_{100}$ and
$N_{\mathrm{H_2O}}/\Delta V$, the averaged 6/4 ratio is consistent with lower
\tdust\ and $f_{75}/f_{100}=1-1.2$. In general, the 6/4 ratio
indicates $T_{\mathrm{dust}}\approx45-80$ K.\footnote{Such high \tdust\ can
be explained in the optically thin case as follows: first, the para-$1_{11}$
level is more easily populated through radiation than the ortho-$2_{12}$
level, because the $B_{lu}/A_{ul}$ ratio for the \t111000\ transition is a
factor 6 higher than for the \t212101\ one ($B_{lu}$ and $A_{ul}$ are the
Einstein coefficients for photo absorption and spontaneous emission). Second, 
the $B_{lu}$ coefficient of the para-\t220111\ pumping transition is a factor
of $\approx2.3$ higher than that of the ortho-\t321212\ pumping
transition. Taking into account an ortho-to-para ratio of 3, a 6/4 ratio of 1
is obtained for $J_{179}J_{75}/(J_{269}J_{101})\approx4.5$ ($J_{\lambda}$ is
the mean specific intensity at wavelength $\lambda$), which requires
$T_{\mathrm{dust}}\approx45$ K. \label{foonote:expl}}
Similarly, the 6/2 ratio is also sensitive to \tdust, as shown in
Fig.~\ref{ratio62}b. The observed averaged 6/2 ratio of 
$\approx1-1.2$ is compatible with \tdust\ somewhat lower than estimated from
the 6/4 ratio. This is attributable to the effects of collisional excitation
of the $2_{02}$ level (thus enhancing line 2 over line 6, see
Fig.~\ref{colis}a and magenta symbols in Fig.~\ref{ratio62}b), or to the
contribution to line 2 by an extended, low \tdust\ component.

There is, however, no observed correlation between the 6/4
ratio and $f_{60}/f_{100}$ (Y13), which should still show a correlation (though
maybe less pronounced) than the expected correlation with $f_{75}/f_{100}$. As
we argue in Sect.~\ref{sec:discussion}, this lack of correlation suggests that
the observed far-IR $f_{60}/f_{100}$ colors, and in particular the observed
$f_{100}$ fluxes, are not dominated by the warm component responsible for the
\hdo\ emission. Indeed, current models for the continuum emission in 
(U)LIRGs indicate that the flux density at 100 $\mu$m is dominated by
relatively cold dust components ($T_{\mathrm{dust}}\sim30$ K)
\citep[e.g.][]{dun03,kov10,cas12}. The observed \hdo\ emission thus arises in
warm regions whose continuum is hidden within the observed far-IR emission,
but may dominate the observed SED at $\lambda\lesssim50$ $\mu$m
\citep[e.g.,][see also Sect.~\ref{sed}]{cas12}.

   \begin{figure}
   \centering
   \includegraphics[width=8.3cm]{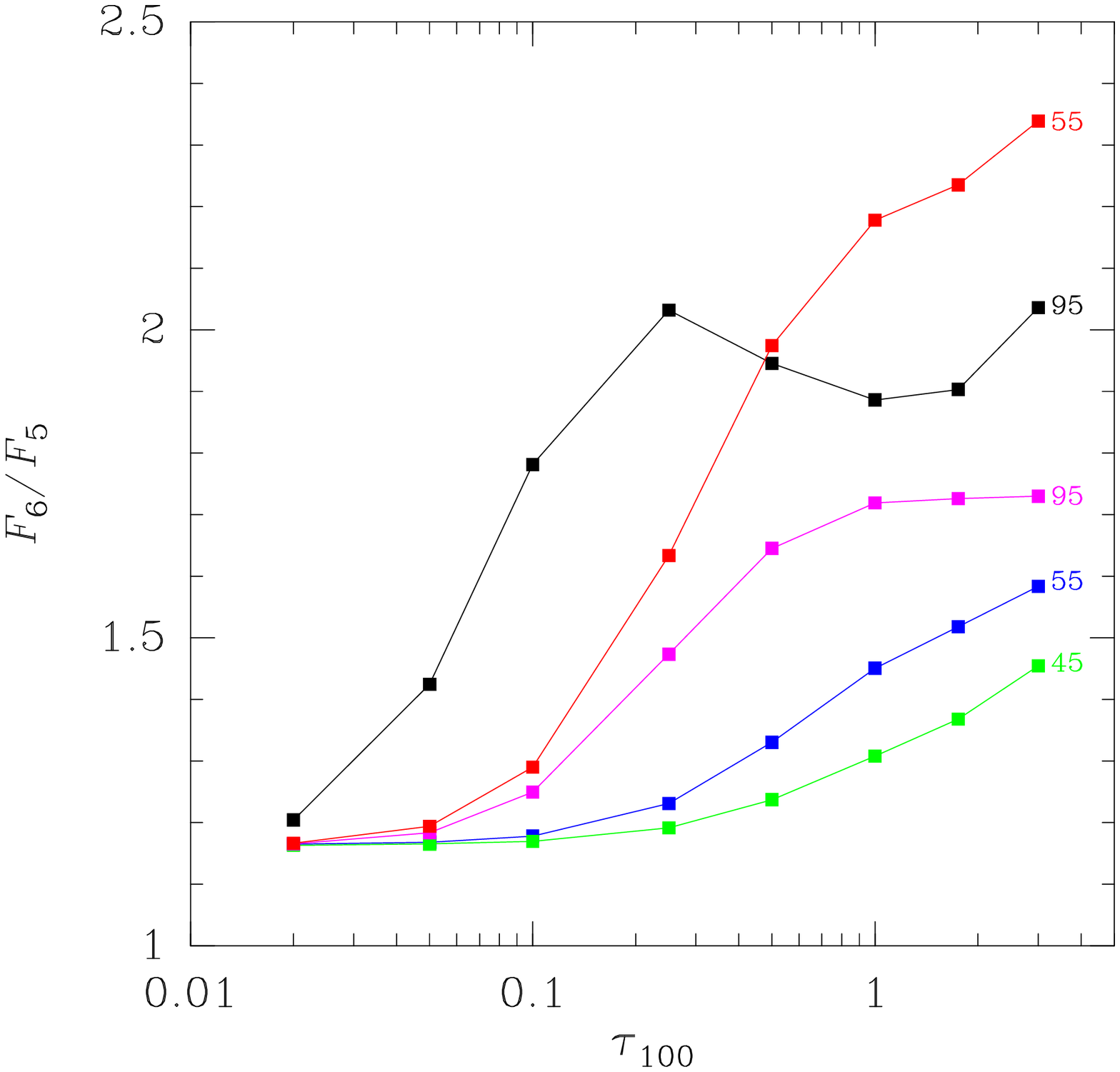}
   \caption{$F_6/F_5$ (o\hdo\ \t321312-to-\t312303) line flux ratio
     as a function of $\tau_{100}$. The small numbers on the right side of the
     curves indicate the values of \tdust\ for each curve. The \hdo\ column
     density per unit of velocity interval is $N_{\mathrm{H_2O}}/\Delta
     V=10^{15}$ $\mathrm{cm^{-2}/(km\,s^{-1})}$ (green, blue, and magenta 
     curves) and $N_{\mathrm{H_2O}}/\Delta V=5\times10^{15}$
     $\mathrm{cm^{-2}/(km\,s^{-1})}$ (red and black curves). 
   }   
    \label{ratio65}
    \end{figure}

The \hdo\ lines 5 and 6 are both pumped through the 75 $\mu$m
transition. Assuming that the lines are optically thin, 
statistical equilibrium of the level populations implies that
every de-excitation in line 6 will be followed by a de-excitation in either
line 5 or in the $3_{12}-2_{21}$ transition (dotted arrow in Fig.~\ref{lev}),
with relative probablities determined by the A-Einstein coefficients. 
In these optically thin conditions, we expect a 6/5
line flux ratio of $F_6/F_5=1.16$ (Fig.~\ref{ratio65}). This is a lower
limit, because in case of high $N_{\mathrm{H_2O}}/\Delta V$ and/or high
\tdust\ and $\tau_{100}$, absorption of line 5 emitted photons that can
eventually be reemited through the $3_{12}-2_{21}$ transition, or absorption
of continuum photons in the \hdo\ \t423312\ transition, will 
decrease the strength of line 5 relative to line 6.  

Although with significant dispersion, overall data for HII+mild AGN sources
indicate $F_6/F_5\sim1.2$ (Y13)\footnote{We infer this value from the
$(L_{\mathrm{H_2O}}/L_{\mathrm{IR}})_2/(L_{\mathrm{H_2O}}/L_{\mathrm{IR}})_i$  
ratios listed in Table~2 by Y13, though $F_6/F_5\sim1.4$ as derived
directly from
$(L_{\mathrm{H_2O}}/L_{\mathrm{IR}})_5/(L_{\mathrm{H_2O}}/L_{\mathrm{IR}})_6$, 
indicating that the averaged $F_6/F_5$ depends on the details (weights) of the
average computation.}, consistent with the optically thin limit;
examples of this galaxy population are NGC~1068 and NGC~6240
\citep{spi12,mei13}. There are, however, sources like Arp~220 and Mrk~231 with
$F_6/F_5\approx1.6$, favoring warm dust ($>55$ K) and substantial columns of
\hdo\ and dust. This indicates that sources in both the optically thin and
optically thick regimes are \hdo\ emitters.

In optically thin conditions and with moderate \tdust, lines $2-4$, together
with 
the pumping \t220111\ 101 $\mu$m transition, form a closed loop (Fig.~\ref{lev})
where statistical equilibrium of the level populations implies equal fluxes
for the three submm lines (Fig.~\ref{mod}a1-c1). 
The rise in \tdust\ and $\tau_{100}$, however,
increases the chance of line absorption in the strong \t322211\
transition at 90 $\mu$m, thus decreasing the flux of line 3 relative to both
line 2 and 4. Consequently, the $F_2/F_3$ ratio is expected to increase from
$\approx1$ (for low $\tau_{100}$) to $\approx2$ (for $\tau_{100}\sim1$ and
$N_{\mathrm{H_2O}}/\Delta V\gtrsim10^{17}$ $\mathrm{cm^{-2}/(km\,s^{-1})}$),
consistent with the relatively high values found in the warm Mrk~231 and
APM~08279 (Y13). If collisional excitation is important (Fig.~\ref{mod}d-f),
$F_2/F_3$ is also expected to increase because collisions mainly boost the
lower lying line 2 (Fig.~\ref{colis}a).

One interesting caveat is, however, the behavior of the 4/3 ratio, because
increasing \tdust\ and/or $N_{\mathrm{H_2O}}$ is predicted to increase
$F_2/F_3$ but maintains $F_4/F_3>1$ (Fig.~\ref{mod}a1-c1). 
In Mrk~231, the high $F_2/F_3$ ratio and mostly the detection of lines $7-8$ 
indicate very warm dust (G-A10), but the relatively low $F_4/F_3\lesssim1$
observed in the source does not match this simple scheme. The problem is
exacerbated with the 6/2 ratio, which is also expected to increase with
increasing \tdust\ and $\tau_{100}$ to $\approx1.5$
(Fig.~\ref{ratio62}), but Mrk~231 shows $F_6/F_2\approx1$. Nevertheless, 
the problem can be solved if source structure is invoked. 
A composite model where a very warm component accounts for the high-lying
lines and a colder (dust) component enhances lines $2-4$ (with probable
contribution from collisionally excited gas, as suggested by the high $F_2/F_3$
ratio), can give a good fit to the SLED (G-A10), although the characteristics
of the ``cold'' component (density, extension, \tdust) are relatively
uncertain. A relatively low flux in line 4 can also be produced by absorption
of continuum photons emanating from a very optically thick component, as in
Arp~220 (see App.~\ref{appa}).

\subsection{The $L_{\mathrm{H_2O}}-L_{\mathrm{IR}}$ correlations} 
\label{sec:discussion}

\subsubsection{H$_2$O and the observed SED} 
\label{sed}

It has long been recognized that single-temperature graybody fits to
galaxy SEDs at far-IR wavelengths often underpredict the observed emission
at $\lambda<50$ $\mu$m. Therefore, multicomponent fitting, based on, for example, a
two-temperature approach, a power-law mass-temperature distribution, a
power-law mass-intensity distribution, or a single cold dust temperature
graybody with a mid-IR power law \citep{dun03,kov10,dal02,cas12}, is required
to match the full SED from the mid-IR to millimeter wavelengths. Our
single-temperature model results on the \hdo\ SLED favors
$T_{\mathrm{dust}}\gtrsim45$ K (Sect.~\ref{lratios}), significantly 
warmer than the cold dust temperatures ($<40$ K) that account for most of
the observed far-IR emission in luminous IR galaxies, indicating that the
\hdo\ submm emission primarily probes the warm region(s) of galaxies where the
mid-IR ($20-50$ $\mu$m) emission is generated (see
footnote~\ref{foonote:expl}).

Relative to the total IR emission of a galaxy, $L_{\mathrm{IR}}^T$, the
contribution to the luminosity by a given \tdust\ component $i$ is
$f_i=L_{\mathrm{IR}}^i/L_{\mathrm{IR}}^T$, and the observed \hdo-to-IR
luminosity ratio is
\begin{equation}
\frac{L_{\mathrm{H_2O}}}{L_{\mathrm{IR}}^T} = 
\sum_i f_i \left( \frac{L_{\mathrm{H_2O}}}{L_{\mathrm{IR}}} \right)_i =
f_{\mathrm{warm}} 
\left( \frac{L_{\mathrm{H_2O}}}{L_{\mathrm{IR}}} \right)_{\mathrm{warm}} +
f_{\mathrm{cold}} 
\left( \frac{L_{\mathrm{H_2O}}}{L_{\mathrm{IR}}} \right)_{\mathrm{cold}}
\end{equation}
where $(L_{\mathrm{H_2O}}/L_{\mathrm{IR}})_i$ are the values plotted in
Fig.~\ref{mod}a2-f2 (for $\Delta V=100$ \kms), and the problem is grossly
simplified by considering only two ``warm'' and ``cold'' components. From the
comparison of the observed average SLED (Y13) with our models, we 
infer that the contribution by the cold component to
$L_{\mathrm{H_2O}}/L_{\mathrm{IR}}^T$ is small, even though $f_{\mathrm{cold}}$
may be high. Since our modeled $L_{\mathrm{IR}}$ emission from the warm 
component is thus only a fraction, $f_{\mathrm{warm}}$, of the total IR budget,
the modeled $L_{\mathrm{H_2O}}/L_{\mathrm{IR}}$ ratios in
Fig.~\ref{mod}a2-f2 should be considered upper limits. The value of
$f_{\mathrm{warm}}$ can only be estimated by fitting the individual
SEDs.

\subsubsection{H$_2$O emission and monochromatic luminosities} 

   \begin{figure}
   \centering
   \includegraphics[width=9cm]{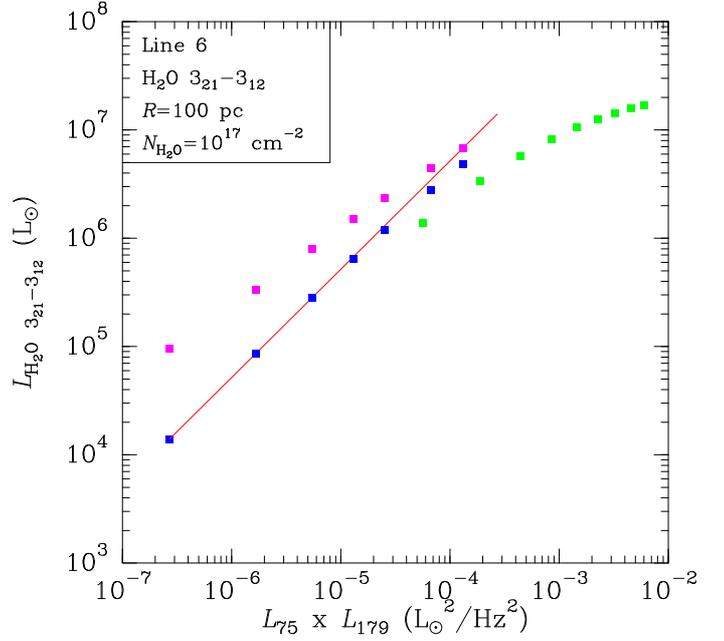}
   \caption{Model results showing the luminosity of the \hdo\ line 6
     (\t321312) as a function of the product of the monochromatic luminosities
     at 75 and 179 $\mu$m. Luminosities are calculated for a source with
     $R_{\mathrm{eff}}=100$ pc, $N_{\mathrm{H_2O}}/\Delta V=10^{15}$
     $\mathrm{cm^{-2}/(km\,s^{-1})}$, and $\Delta V=100$ \kms. 
     Blue squares indicate models 
     with $\tau_{100}=0.1$, resulting in optically thin or
     moderately thick \hdo\ emission, without collisional excitation. Magenta
     squares show results for the same models but with collisional excitation
     included with $T_{\mathrm{gas}}=150$ K and
     $n_{\mathrm{H_2}}=3\times10^{5}$ \cmt. Green symbols indicate models with
     $\tau_{100}=1.0$, resulting in optically thick \hdo\ emission. For
     optically thin \hdo\ emission and without collisional excitation, the
     models indicate a linear correlation 
     between $L_{\mathrm{H_2O}}$ and $L_{75}\times L_{179}$ (red line).
   }   
    \label{lh2o6}
    \end{figure}

The \hdo\ submm emission of lines $2-6$ essentially involves two excitation
processes, that of the base level ($2_{12}$ for ortho and $1_{11}$ for
para-\hdo) and absorption in the transitions at 75 $\mu$m (ortho) or 101
$\mu$m (para, Fig.~\ref{lev}). If collisional excitation is unimportant, the
excitation of the base levels is also produced by absorption of dust-emitted
photons in the corresponding transitions, i.e., in the $\t212101$ line at 179
$\mu$m (ortho) or $\t111000$ at 269 $\mu$m (para). In optically thin
conditions and for fixed $N_{\mathrm{H_2O}}$ and $\Delta V$, our models then
show a linear correlation between the \hdo\ luminosities $L_{\mathrm{H_2O}}$
and the product of the continuum monochromatic luminosities responsible for
the excitation, $L_{179}\times L_{75}$ (ortho) or $L_{269}\times L_{101}$
(para). This linear correlation is illustrated in Fig.~\ref{lh2o6} for line
6. The linear correlation, however, breaks down when the line becomes
optically thick or when collisional excitation becomes important (in which
case, $L_{\mathrm{H_2O}}$ is independent of $L_{179(269)}$).

\subsubsection{The $L_{\mathrm{H_2O}}/L_{\mathrm{IR}}$ ratios and
  $T_{\mathrm{dust}}$}  
\label{sec:lh2olir-tdust}

The above considerations are relevant for our understanding of the behavior of 
the modeled $L_{\mathrm{H_2O}}/L_{\mathrm{IR}}$ values with variations
in \tdust. In the optically thin case and with collisional excitation ignored,
the double dependence of $L_{\mathrm{H_2O}}$ on two monochromatic luminosities
implies that $L_{\mathrm{H_2O}}$ is (nearly) proportional to $L_{\mathrm{IR}}$. Our
predicted SEDs indicate that, for small variations in \tdust\ around $55$ K,
$L_{269}\propto L_{\mathrm{IR}}^{2/7}$ and $L_{101}\propto
L_{\mathrm{IR}}^{1/2}$. Therefore, for the
para-\hdo\ lines $2-4$, $L_{2-4}\propto L_{\mathrm{IR}}^{0.8}$ in optically
thin conditions, slightly slower than linear. For the ortho lines,
$L_{179}\propto L_{\mathrm{IR}}^{1/3}$ and $L_{75}\propto
L_{\mathrm{IR}}^{2/3}$, so that $L_{5-6}\propto L_{\mathrm{IR}}$. This
explains why, in Fig.~\ref{mod}(a2-b2), the
$L_{\mathrm{2-4}}/L_{\mathrm{IR}}$ ratios show a slight 
decrease with increasing \tdust\ above 55 K, while
$L_{\mathrm{5-6}}/L_{\mathrm{IR}}$ versus \tdust\ attain a maximum at
$T_{\mathrm{dust}}\approx55$ K in optically thin models that omit 
collisional excitation. These results are
robust against variations in the spectral index of dust down to $\beta=1.5$
(Sect.~\ref{sec:kabs}). 

   \begin{figure}
   \centering
   \includegraphics[width=9cm]{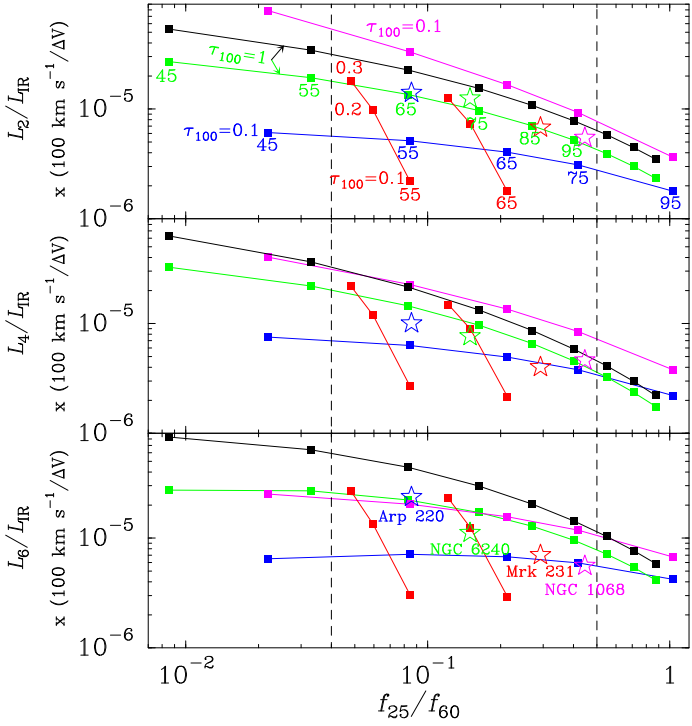}
   \caption{Modeled $L_{\mathrm{H_2O}}/L_{\mathrm{IR}}\times (100 \,
     \mathrm{km\,s^{-1}}/\Delta V)$ for lines 
     2 (upper), 4 (middle), and 6 (lower) as a function of the $f_{25}/f_{60}$
     color. The dashed vertical lines indicate the lower and upper
     limits for $f_{25}/f_{60}$ measured by Y13. In the upper panel, 
     the small numbers below the squares 
     indicate the value of \tdust, and $\tau_{100}$ is also indicated.
     Blue squares: $N_{\mathrm{H_2O}}/\Delta V=10^{15}$
     $\mathrm{cm^{-2}/(km\,s^{-1})}$ and $\tau_{100}=0.1$; 
     magenta: same as blue symbols but with collisional excitation included with
     $T_{\mathrm{gas}}=150$ K and $n_{\mathrm{H_2}}=3\times10^{5}$ \cmt; green:
     $N_{\mathrm{H_2O}}/\Delta V=10^{15}$ $\mathrm{cm^{-2}/(km\,s^{-1})}$ and
     $\tau_{100}=1.0$; black: same as green but with $N_{\mathrm{H_2O}}/\Delta
     V=5\times10^{15}$ $\mathrm{cm^{-2}/(km\,s^{-1})}$. Red squares show 
     results for fixed $T_{\mathrm{dust}}=55$ and $65$ K with
     $N_{\mathrm{H_2O}}/(\Delta V\tau_{100})=5\times10^{15}$
     $\mathrm{cm^{-2}/(km\,s^{-1})}$ and $\tau_{100}=0.1-0.3$.  
     The starred symbols indicate the positions of Arp~220 (blue),
     NGC~6240 (green), Mrk~231 (red), and NGC~1068 (magenta), as reported by
     \cite{ran11}, \cite{mei13}, G-A10, and Y13, respectively. When
     compared with observations, the modeled $L_{\mathrm{IR}}$ 
     values should be considered a fraction of the observed IR luminosities 
     (Sect.~\ref{sed}).
     If $f_{60}$ is contaminated by cold dust, the points would move to the
     left. 
   }   
    \label{f25f60}
    \end{figure}

In Fig.~\ref{f25f60} we show the $L_{\mathrm{H_2O}}/L_{\mathrm{IR}}$ ratios
(with $\Delta V=100$ \kms; $L_{\mathrm{H_2O}}/L_{\mathrm{IR}}\propto \Delta
V$) for lines 2, 4, and 6 as a function of the $f_{25}/f_{60}$ color. The
observed $f_{25}/f_{60}$ was used by Y13 to characterize the \hdo\
emission and is especially relevant given that the \hdo\ submm emission
arises in warm regions in which the mid-IR continuum emission is not severely
contaminated by cold dust. However, the continuum at $\lambda=60$ $\mu$m may
still be contaminated to some extent, in which case the data points in
Fig.~\ref{f25f60} will move toward the left. We also recall that the
$L_{\mathrm{H_2O}}/L_{\mathrm{IR}}$ values are upper limits.

The first conclusion inferred from Fig.~\ref{f25f60} is that the range of 
$f_{25}/f_{60}$ colors measured by Y13 (between the dashed lines) matches
\tdust\ in the ranges favored by the observed \hdo\ line flux ratios,
that is, $50-75$ K and optically thin conditions ($\tau_{100}=0.1$) and also $T_{\mathrm{dust}}=60-95$ K and $\tau_{100}=1.0$.
 This indicates that the warm environments
responsible for the \hdo\ emission are best traced in the continuum in this
wavelength range, but also that the $f_{25}/f_{60}$ color alone 
involves degeneracy in the dominant \tdust\ and $\tau_{100}$ responsible for
the mid-IR continuum emission. 
As shown in Sect.~\ref{lratios}, the first set of conditions can explain the
line ratios $2-6$ in warm sources (where lines $7-8$ are not detected to a
significant level), while the second set is required to explain the
\hdo\ emission in very warm sources (with detection of lines $7-8$).

Second, it is also relevant that the $L_{\mathrm{H_2O}}/L_{\mathrm{IR}}$
values differ by a factor $\lesssim2$ between models with warm dust in the
optically thin regime ($T_{\mathrm{dust}}=55$ K, $\tau_{100}=0.1$,
$N_{\mathrm{H_2O}}/\Delta V=10^{15}$ $\mathrm{cm^{-2}/(km\,s^{-1})}$) and
those with very warm dust in the 
optically thick regime with high \hdo\ columns ($T_{\mathrm{dust}}=95$ K,
$\tau_{100}=1$, $N_{\mathrm{H_2O}}/\Delta V=5\times10^{15}$
$\mathrm{cm^{-2}/(km\,s^{-1})}$), potentially 
explaining why sources with different physical conditions show similar
$L_{\mathrm{H_2O}}/L_{\mathrm{IR}}$ ratios (Y13).

Third, in optically thin conditions ($\tau_{100}\sim0.1$) and if
collisional excitation is unimportant, the models with constant
$N_{\mathrm{H_2O}}/\Delta V=10^{15}$ $\mathrm{cm^{-2}/(km\,s^{-1})}$ (blue
symbols) predict a slow decrease in $L_{2-4}/L_{\mathrm{IR}}$ and a nearly 
constant $L_{5-6}/L_{\mathrm{IR}}$ with increasing $f_{25}/f_{60}$, as argued
above. This behavior, however, fails to match the observed trends (Y13), 
as $L_{2-4}/L_{\mathrm{IR}}$ and $L_{6}/L_{\mathrm{IR}}$ decrease by factors of
$\sim2$ and $\sim3$, respectively, when $f_{25}/f_{60}$ 
increases from $\lesssim0.08$ to $\gtrsim0.15$. When collisional excitation
is included (magenta symbols), the $L_{2-4}/L_{\mathrm{IR}}$ ratios show a
stronger dependence on $f_{25}/f_{60}$, but $L_{6}/L_{\mathrm{IR}}$
still changes only slightly with $f_{25}/f_{60}$. 

Therefore, optically thin models with varying \tdust\ but constant
$\tau_{100}$, $N_{\mathrm{H_2O}}/\Delta V$, and $\Delta V$ cannot account for
the observed $L_{\mathrm{H_2O}}/L_{\mathrm{IR}}-f_{25}/f_{60}$ trend. This
indicates that, in optically thin galaxies, parameters other than \tdust\ are 
systematically varied when $f_{25}/f_{60}$ is increased and that
optically thick sources also contribute to the observed trend: \\ 
($i$) Galaxies in the optically thin regime (with $\tau_{100}<1$) are
predicted to show a very steep dependence of $L_{\mathrm{H_2O}}/L_{\mathrm{IR}}$
on $\tau_{100}$ for constant \tdust\ and $N_{\mathrm{H_2O}}/(\Delta V\tau_{100})$
(that is, for constant \hdo\ abundance), with higher $\tau_{100}$ impling {\it
  lower} $f_{25}/f_{60}$. We illustrate this point in Fig.~\ref{f25f60} with
the red squares, corresponding to fix 
$T_{\mathrm{dust}}=55$ and $65$ K and $N_{\mathrm{H_2O}}/(\Delta
V\tau_{100})=5\times10^{15}$ $\mathrm{cm^{-2}/(km\,s^{-1})}$,
with $\tau_{100}$ ranging from $0.1$ to $0.3$. Therefore, we
expect that the observed increase in $f_{25}/f_{60}$ is not only due to
an increase in \tdust\ from source to source, but also to variations in
$\tau_{100}$ in the optically thin regime. Examples of galaxies in this regime
are the AGNs NGC~6240 and NGC~1068 (see also App.~\ref{appa}). \\
($ii$) In the optically thick regime ($\tau_{100}\gtrsim1$), galaxies are also
predicted to show a relatively steep variation in
$L_{\mathrm{H_2O}}/L_{\mathrm{IR}}$ with $f_{25}/f_{60}$ due to increasing
\tdust\ (black symbols in Fig.~\ref{f25f60}) because  the \hdo\ lines
saturate and their luminosities flatten with increasing monochromatic
luminosities (Fig.~\ref{lh2o6}). Extreme examples of this
galaxy population are Arp~220 and Mrk~231. Line saturation also implies that
the $L_{\mathrm{H_2O}}/L_{\mathrm{IR}}$ ratios are not much higher than in the
optically thin case even if much higher $N_{\mathrm{H_2O}}/\Delta
V=5\times10^{15}$ $\mathrm{cm^{-2}/(km\,s^{-1})}$
are present, and the corresponding ratios are consistent with the
observed values to within the uncertainties in $f_{\mathrm{warm}}$. The
presence of even warmer dust ($>100$ K) with significant contribution to
$L_{\mathrm{IR}}$ will additionally decrease
$L_{\mathrm{H_2O}}/L_{\mathrm{IR}}$ (Y13). 

In summary, the steep decrease in $L_{\mathrm{H_2O}}/L_{\mathrm{IR}}$ at
$f_{25}/f_{60}\approx0.1-0.15$ measured by Y13 is
consistent with both types of galaxies (with optically thin and optically
thick continuum) populating the diagram and suggests that the observed
variations in $f_{25}/f_{60}$ are not only due to variations in \tdust\ but
also to variations in $\tau_{100}$ in the optically thin regime. At the other
extreme, the optically thick (saturated) and very warm galaxies are also
expected to show a decrease in $L_{\mathrm{H_2O}}/L_{\mathrm{IR}}$ with
increasing \tdust\ (and $f_{25}/f_{60}$), as anticipated by Y13. To
distinguish between both regimes for a given galaxy, the line ratios
(specifically $F_6/F_5$, Sect.~\ref{lratios}) and mostly the detection of
lines $7-8$ or the detection of high-lying \hdo\ absorption lines at
far-IR wavelengths are required. 
The observations reported by Y13 indicate that these optically thick and warm
components (diagnosed by the detection of lines $7-8$) are present in at least
ten sources. At least in NGC~1068 the upper limits on lines $7-8$ are
stringent (S12), allowing us to infer optically thin conditions.

\subsubsection{Line saturation and a theoretical upper limit to
  $L_{\mathrm{H_2O}}/L_{\mathrm{IR}}$} 

   \begin{figure}
   \centering
   \includegraphics[width=9cm]{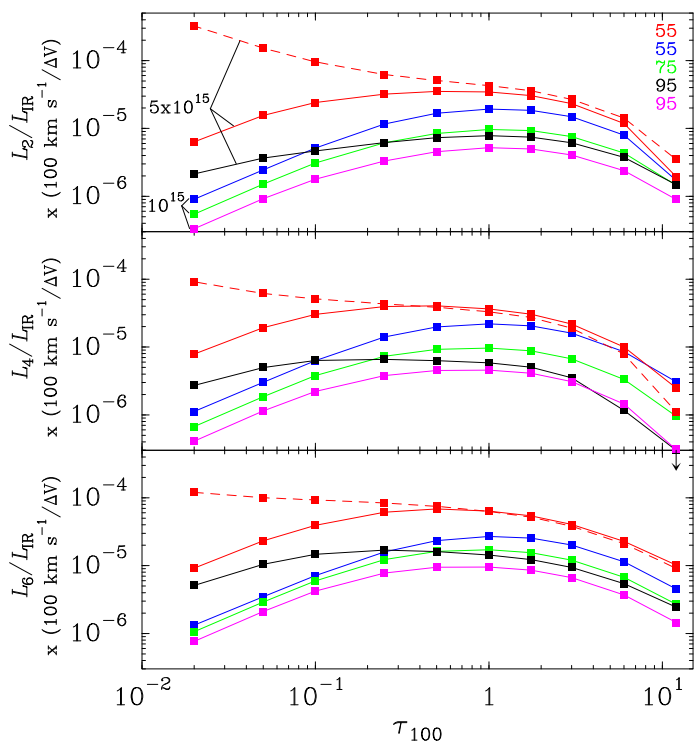}
   \caption{Modeled $L_{\mathrm{H_2O}}/L_{\mathrm{IR}}\times (100 \,
     \mathrm{km\,s^{-1}}/\Delta V)$ values for lines 
     2 (upper), 4 (middle), and 6 (lower) as a function of
     $\tau_{100}$. Collisional excitation is
     ignored except for the red dashed lines, where $T_{\mathrm{gas}}=150$ K
     and $n_{\mathrm{H_2}}=3\times10^{5}$ \cmt\ are adopted. The small numbers
     on the right side of the upper panel indicate 
     the value of \tdust\ in K, and those on the left side indicate
     $N_{\mathrm{H_2O}}/\Delta V$ in $\mathrm{cm^{-2}/(km\,s^{-1})}$. When 
     compared with observations, the modeled $L_{\mathrm{IR}}$ 
     values should be considered a fraction of the observed IR luminosities
     (Sect.~\ref{sed}). For $\tau_{100}=12$ and 
     $T_{\mathrm{dust}}=95$ K, line 4 is 
     predicted to be in absorption. For fixed \tdust\ and
       $\tau_{100}\sim1$, the $L_{\mathrm{H_2O}}/L_{\mathrm{IR}}$ ratios are
       similar for very different $N_{\mathrm{H_2O}}$, indicating line 
       saturation.
   }   
    \label{tau100}
    \end{figure}

Saturation of the \hdo\ submm lines in optically thick
  ($\tau_{100}\sim1$) sources implies that there is an upper limit on
  $L_{\mathrm{H_2O}}/L_{\mathrm{IR}}\times (100 \, \mathrm{km\,s^{-1}}/\Delta
  V)$  that, in the absence of significant collisional excitation, cannot be
  exceeded. In Fig.~\ref{tau100}, the $L_{\mathrm{H_2O}}/L_{\mathrm{IR}}\times
  (100 \, \mathrm{km\,s^{-1}}/\Delta V)$ ratios for lines 
  2, 4, and 6 are plotted as a function of $\tau_{100}$ for the most
  favored \tdust\ range of $55-95$ K and
  $N_{\mathrm{H_2O}}/\Delta V=(1-5)\times10^{15}$
  $\mathrm{cm^{-2}/(km\,s^{-1})}$. In optically thin conditions 
  ($\tau_{100}\lesssim0.1$ for $N_{\mathrm{H_2O}}/\Delta V=10^{15}$
  $\mathrm{cm^{-2}/(km\,s^{-1})}$) and without 
  collisions, $L_{\mathrm{H_2O}}/L_{\mathrm{IR}}$ scales linearly (for fixed
  $\Delta V$) with $\tau_{100}$ because $L_{179(269)}\times
  L_{75(101)}\propto\tau_{100}^2$ while 
  $L_{\mathrm{IR}}\propto\tau_{100}$. The curves flatten as the \hdo\ lines
  saturate and show a maximum at $\tau_{100}\approx0.5-1$. Values of
  $\tau_{100}$ significantly higher than unity are predicted to decrease
  $L_{\mathrm{H_2O}}/L_{\mathrm{IR}}$. In very optically thick components
  of very warm sources, the submm lines are predicted to be observed in weak
  emission or even in absorption, especially in line 4. Arp~220
  is a case in point \citep{sak08}, in which the \hdo\ submm emission is
  expected to arise from a region that surrounds the optically thick
  nuclei (see App.~\ref{appa}). For $\Delta V=100$ \kms, the maximum
  attainable values of $L_{\mathrm{H_2O}}/L_{\mathrm{IR}}$ (red curves) are
  $3.5\times10^{-5}$, $4\times10^{-5}$, and $7\times10^{-5}$ for lines 2, 4,
  and 6, respectively, comfortably higher than the values observed in any
  source by Y13. Recently, a value of $L_{2}/L_{\mathrm{IR}}=(4.3\pm
  1.6)\times10^{-5}$ has been measured in the submillimeter galaxy SPT
  0538-50, a gravitationally lensed dusty star-forming galaxy at $z \approx
  2.8$ \citep{both13}. Although the authors do not exclude differential
  lensing effects, which could affect the line-to-luminosity ratios, this
  value is still consistent with our upper limit, 
  suggesting strong saturation in this source. In HFLS3 at $z=6.34$,
  \cite{rie13} have measured $F_6/F_2=2.2\pm0.5$ and 
  $F_6/F_5=2.6\pm0.8$; within the uncertainties, these values are consistent
  with warm or very warm $T_{\mathrm{dust}}\gtrsim65$ K and
  high $N_{\mathrm{H_2O}}$ (Figs.~\ref{ratio62}-\ref{ratio65}). The \hdo\
  lines are most likely saturated in HFLS3 as is also 
  indicated by the $L_{6}/L_{\mathrm{IR}}=(7.7\pm1.3)\times10^{-5}$ ratio,
  which is still consistent with the strong saturation limit for warm \tdust\
  given the very broad linewidth of the \hdo\ line ($\sim940$
  \kms; see Sect~\ref{sec:param}). O13 reported
  $L_{2}/L_{\mathrm{IR}}=(0.5-2)\times10^{-5}$ in 
    high-$z$ ultra-luminous infrared galaxies, also consistent with the upper
    limit in Fig.~\ref{tau100} even for $T_{\mathrm{dust}}\sim75$ K when
    taking the 
    broad line widths of the \hdo\ lines into account. Line saturation and 
a relatively small contribution from cold dust to the infrared emission in
these extreme galaxies are implied. With collisional excitation in optically
thin environments with moderate \tdust\ but high $N_{\mathrm{H_2O}}$, 
the above $L_{\mathrm{H_2O}}/L_{\mathrm{IR}}$  ratios (red dashed lines in
Fig.~\ref{tau100}) may even attain higher values, though the adopted
$\Delta V=100$ \kms\ is too high for $\tau_{100}<0.3$ and
$n_{\mathrm{H_2}}=3\times10^{5}$ \cmt\ (Sect~\ref{sec:param},
eq.~\ref{eq:deltavvir}).

\subsubsection{The correlation}
\label{sec:corr}

   \begin{figure}
   \centering
   \includegraphics[width=8cm]{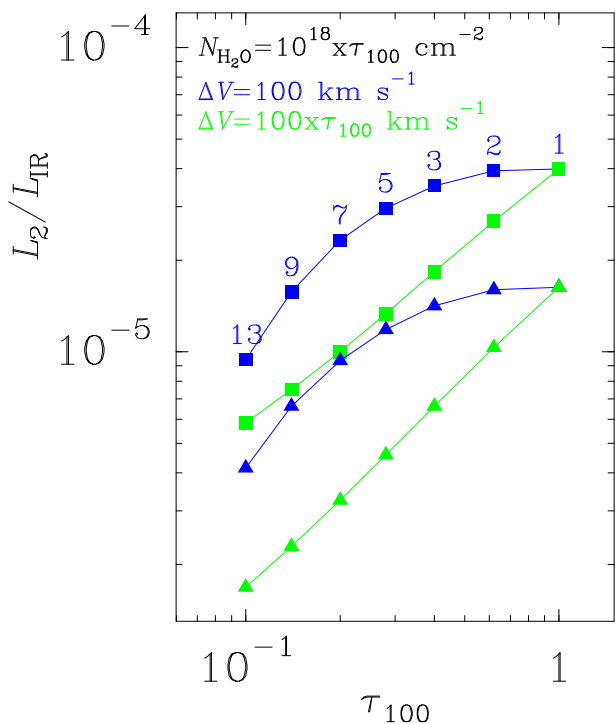}
   \caption{Modeled $L_{\mathrm{2}}/L_{\mathrm{IR}}$ ratio  
     as a function of $\tau_{100}$. Squares and triangles indicate
     $T_{\mathrm{dust}}=55$ and $75$ K, respectively. In all models, 
     collisional excitation is included with
     $T_{\mathrm{gas}}=150$ K and $n_{\mathrm{H_2}}=3\times10^{4}$ \cmt. 
     $N_{\mathrm{H_2O}}/\tau_{100}=10^{18}$ \cmd\ is adopted, corresponding to
     a constant \hdo\ abundance of $7.7\times10^{-7}$ (Eq.~\ref{eq:nh}).
     Blue symbols indicate models with $\Delta V=100$ \kms\ and thus with
     variable $K_{\mathrm{vir}}$ (Eq.~\ref{eq:kvir}) indicated with the 
     numbers. Green symbols show results with 
     $\Delta V=100\times\tau_{100}$ \kms\ simulating a constant value of
     $K_{\mathrm{vir}}=1.3$. When 
       compared with observations, the modeled $L_{\mathrm{IR}}$ values should
       be considered a fraction of the observed IR luminosities 
       (Sect.~\ref{sed}), and thus the 
       modeled $L_{\mathrm{2}}/L_{\mathrm{IR}}$ values are upper limits.
   }   
    \label{tau100b}
    \end{figure}

The broad range in observed 
$L_{\mathrm{IR}}$ in luminous IR galaxies with \hdo\ emission
may be attributable to varying the effective size of the emitting region.
As noted in Sect.~\ref{sec:param}, varying $R_{\mathrm{eff}}$ (equivalent to
varying the number of individual 
regions that contribute to $L_{\mathrm{IR}}$ or to increasing
$L_{\mathrm{IR}}$ for a single source) is expected to generate linear
$L_{\mathrm{H_2O}}-L_{\mathrm{IR}}$ correlations if the other parameters 
(\tdust, $\tau_{100}$, \tgas, $n_{\mathrm{H_2}}$, $N_{\mathrm{H_2O}}/\Delta
V$, and $\Delta V$) remain constant. 

In Fig.~\ref{tau100b} we show the $L_2/L_{\mathrm{IR}}$ ratio as a
  function of $\tau_{100}$ for models with $T_{\mathrm{dust}}=55$ K
  and $T_{\mathrm{dust}}=75$ K that 
  assume a constant \hdo-to-dust opacity ratio, that is,
  $N_{\mathrm{H_2O}}/\tau_{100}=10^{18}$ \cmd. According to Eq.~(\ref{eq:nh}),
  this corresponds to a constant \hdo\ abundance of $7.7\times10^{-7}$. Both
  models with $\Delta V=100$ \kms\ (independent of $\tau_{100}$), and $\Delta
  V/\tau_{100}=100$ \kms\ (corresponding to a constant 
  $K_{\mathrm{vir}}=1.3$) are shown. The figure 
  illustrates that a supralinear correlation between $L_{\mathrm{H_2O}}$ and
  $L_{\mathrm{IR}}$ can be expected if, on average, $\tau_{100}$ is an increasing
  function of $L_{\mathrm{IR}}$. If most sources with
  $L_{\mathrm{IR}}\sim5\times10^{10}$ \Lsun\ were optically thin
  ($\tau_{100}\sim0.1$), and the high-$z$ sources with
  $L_{\mathrm{IR}}\sim10^{13}$ \Lsun\ (O13)
  were mostly optically thick ($\tau_{100}\sim1$), one would then expect
  $L_2\propto L_{\mathrm{IR}}^{1.3}$ from Fig.~\ref{tau100b}, which can
  account for the observed supralinear correlation found by O13 and
  Y13. However, similar supralinear correlations would then be expected for
  the other submm lines $3-6$.

\section{Summary of the model results for optically 
classified starbursts and AGNs} 
\label{sec:summary}

Following the optical classification of sources by Y13 into optically
classified star-formation-dominated galaxies with possible mild AGN
contribution (HII+mild AGN sources) and optically identified strong-AGN
sources, we now consider these two groups separately.

\subsection{HII+mild AGN sources}

\begin{figure}
   \centering
   \includegraphics[width=9cm]{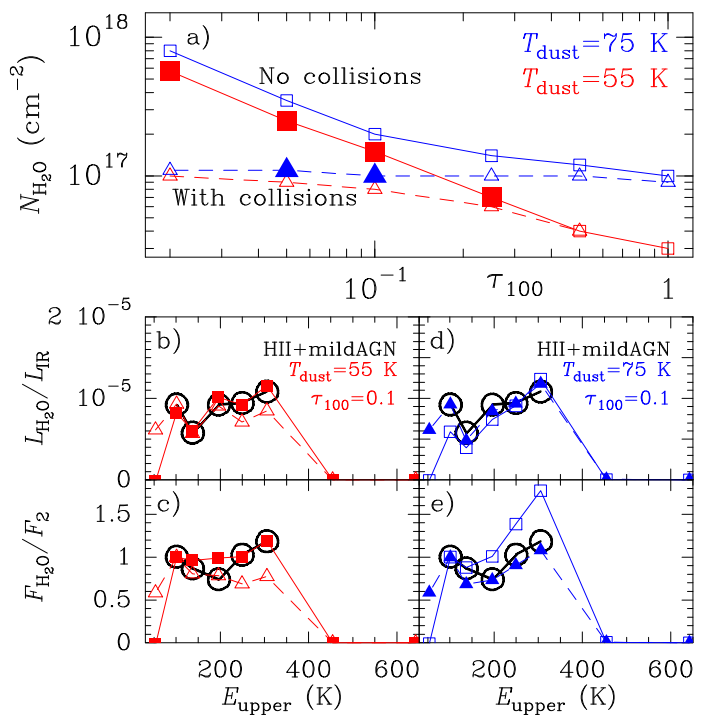}
   \caption{{\bf a)} Values of $N_{\mathrm{H_2O}}$ for $\Delta V=100\,
     \mathrm{km\,s^{-1}}$ as a function of 
     $\tau_{100}$, for $T_{\mathrm{dust}}=55$ K (red) and $75$ K (blue),
     required to account for the observed averaged
     $L_{\mathrm{H_2O}}/L_{\mathrm{IR}}$ ratios in HII-mild AGN sources (as
     given by Y13). Models both without (squares) and with (triangles)
       collisional excitation are shown. In the latter models,
     $T_{\mathrm{gas}}=150$ K, and $n_{\mathrm{H_2}}=5\times10^4-3\times10^5$
     \cmt\ for $T_{\mathrm{dust}}=55-75$ K, respectively. The large filled
     symbols indicate the best fit models to the averaged SLED. 
     Panels {\bf b-e)} compare in detail the four models with $\tau_{100}=0.1$
     (both with and without collisions) with the observed values 
     (black circles) in
     HII-mild AGN sources (both the normalized flux ratios (SLED) and the 
     $L_{\mathrm{H_2O}}/L_{\mathrm{IR}}$ ratios, Y13, for lines 2-6). In
     panels {\bf b)} and {\bf d)}, the models indicate the values of 
       $L_{\mathrm{H_2O}}/L_{\mathrm{IR}}$ for $\Delta V=100$ \kms.
     Lines 1, 7, and 8 are excluded from the comparison because of their low
     detection rates (Y13).
   }   
    \label{best}
    \end{figure}

We focus here on those HII+mild AGN sources where lines $2-6$
are detected but lines $7-8$ are undetected (that is, ``warm'' sources as
defined in Sect.~\ref{sec:general}). The average \hdo\ flux ratios 
reported by Y13 (their Table~2) indicate that $(i)$ $F_6/F_2\sim1.2$,
favoring $T_{\mathrm{dust}}=55$ K if there is no significant collisional
excitation and $T_{\mathrm{dust}}=75$ K if the \hdo\ emission arises in warm
and dense gas (Fig.~\ref{ratio62}); $(ii)$ $F_6/F_5\sim1.2$, consistent with
the optically thin regime (Fig.~\ref{ratio65}).  
For these \tdust, Fig.~\ref{best}a shows the values of
$N_{\mathrm{H_2O}}$ for $\Delta V=100$ \kms\ required to
explain the observed $L_{\mathrm{H_2O}}/L_{\mathrm{IR}}$ ratios, 
as a function of $\tau_{100}$. Models with included or excluded collisional
excitation are considered. We recall that $\Delta V$ is the velocity
dispersion of the dominant structure(s) that accounts for the \hdo\ emission
(Sect.~\ref{sec:param}), and for the case of low $\tau_{100}$ and relatively
high densities, Eq.~(\ref{eq:deltavvir}) suggests $\Delta V<100$ \kms\ with the
consequent increase in $N_{\mathrm{H_2O}}$ (Fig.~\ref{tau100b}).

The decrease in $\tau_{100}$ implies the increase in
$N_{\mathrm{H_2O}}$ in optically thin conditions and when collisional excitation
is unimportant. Our best fit models for the average SLED (big solid symbols)
favor optically thin far-IR emission ($\tau_{100}\lesssim0.3$). In
Fig.~\ref{best}b-e, the detailed comparison between the $\tau_{100}=0.1$ models
and the observations (Y13) is shown. Significant collisional excitation is not
favored for $T_{\mathrm{dust}}=55$ K, since it would increase $F_2$ relative to
$F_6$. In addition, these optically thin models have the drawback of
overestimating $F_4/F_2$. Conversely, the $T_{\mathrm{dust}}=75$ K models
favor significant collisional excitation in order to increase $F_2$ relative
to $F_6$. The very optically thin models ($\tau_{100}\lesssim0.05$) are also
 not favored given the very high amounts of \hdo\ required to explain (with no
collisional excitation) the $L_{\mathrm{H_2O}}/L_{\mathrm{IR}}$ ratios.

In summary, $T_{\mathrm{dust}}=55-75$ K, $\tau_{100}\sim0.1$, and
$N_{\mathrm{H_2O}}\sim(0.5-2)\times10^{17}$ \cmd\ can explain the bulk of the 
\hdo\ submm emission in warm star-forming galaxies
  (Table~\ref{tab:par}). As shown in Fig.~\ref{f25f60},
$T_{\mathrm{dust}}=55$ K and $\tau_{100}=0.1-0.2$ predict 25-to-60 $\mu$m flux
density ratios of $f_{25}/f_{60}=(8.5-6.0)\times10^{-2}$, in agreement with 
the observed values for the bulk of sources (Y13), while
$T_{\mathrm{dust}}=75$ K and $\tau_{100}=0.1-0.2$ predict
$f_{25}/f_{60}=0.42-0.30$ (close to the observed upper values). 
Assuming a gas-to-dust ratio of 100 by mass, $\tau_{100}\sim0.1$
corresponds to a column density of H nuclei of
$N_{\mathrm{H}}\approx1.3\times10^{23}$ \cmd\ (Eq.~\ref{eq:nh}), and thus an
\hdo\ abundance of $X_{\mathrm{H_2O}}\sim10^{-6}$. To within a factor of 3 
uncertainty due to the $\tau_{100}-N_{\mathrm{H}}$ calibration, the specific
values used for $\tau_{100}$ and $\Delta V$, and variations in the
measurements for individual sources, this is the typical \hdo\ 
abundance that we infer from the observed $L_{\mathrm{H_2O}}-L_{\mathrm{IR}}$
correlation. Molecular shocks and hot core chemistry are very likely responsible
for this $X_{\mathrm{H_2O}}$, which is well above the volume-averaged values
inferred in Galactic dark clouds and PDRs
\citep[e.g.,][]{ber00,sne00,mel05,dis11}.

Finally, we note that $T_{\mathrm{dust}}=55$ K and $\tau_{100}=0.1-0.2$, 
and the assumption that most of the IR is powered by star formation in
these sources of Y13, imply a star-formation-rate surface density\footnote{
    $\Sigma_{\mathrm{SFR}}$ is estimated as $10^{-10}L_{\mathrm{IR}}/(\pi
    R^2)$, where a \cite{cha03} initial mass function is used, and
    $\Sigma_{\mathrm{gas}}$ is given by 
    $M_{\mathrm{gas}}/(\pi R^2)$ where $M_{\mathrm{gas}}= \frac{4\pi}{3}
    N_{\mathrm{H}}m_{\mathrm{H}}R^2$ with
    $N_{\mathrm{H}}=1.3\times10^{24}\tau_{100}$ \cmd\ (eq.~\ref{eq:nh}).}
of $\Sigma_{\mathrm{SFR}}=121-195$ \Msun\ yr$^{-1}$
kpc$^{-2}$ and gas mass surface density of
$\Sigma_{\mathrm{gas}}=1430-2860$ \Msun\ pc$^{-2}$.  
The implied depletion or exhaustion time scale,
  $t_{\mathrm{dep}}=\Sigma_{\mathrm{gas}}/\Sigma_{\mathrm{SFR}}$, is
$\sim12-15$ Myr. These values lie close to the
$\Sigma_{\mathrm{SFR}}-\Sigma_{\mathrm{dense}}$ star-formation correlation
found by \cite{gbu12} from HCN emission in (U)LIRGs with their
revised HCN-$M_{\mathrm{dense}}$ conversion factors. This agreement
suggests that the submm \hdo\ and the mm HCN emission in
(U)LIRGs arise from the same regions. Sources with $T_{\mathrm{dust}}=75$ K
would imply even shorter time scales and suggest high rates of ISM return from
SNe and stellar winds. A follow-up study of the 
relationship between $L_{\mathrm{H_2O}}$ and $L_{\mathrm{HCN}}$ is required 
to check this point. In addition, modeling the individual sources 
simultaneously in the continuum and the \hdo\ emission  
will provide further constraints on the nature of these regions.

\subsection{Strong optically classified AGN sources}

The general finding that the \hdo\ emission is similar in star-forming
  and strong-AGN sources (Y13) may simply indicate that the far-IR pumping of
  \hdo\ occurs regardless of whether the dust is heated via star
    formation or an AGN. There are, however, some differences between
  the two source types. Strong AGNs show a higher
  detection rate in \hdo\ \t111000\ 
  (Y13), indicating that the gas densities are higher in the
  circumnuclear regions of AGNs. Another difference is that the
  $L_{\mathrm{H_2O}}/L_{\mathrm{IR}}$ ratios are somewhat lower in strong AGN
  sources (Y13). While relatively low columns of dust and \hdo\ in these
    sources could explain this observational result, it is also possible that
    high X-ray fluxes photodissociate \hdo,\ reducing its abundance relative to
    star-forming galaxies. High 
  abundances of \hdo\ require effective shielding from UV and X-ray photons  and thus high columns of dust and gas that, in AGN-dominated galaxies,
    may be effectively provided by an optically thick torus probably
    accompanied by starburst activity. In addition, warm dust further 
  enhances $X_{\mathrm{H_2O}}$ through an undepleted chemistry and pumps the
  excited \hdo\ levels, while warm gas
  will further boost $X_{\mathrm{H_2O}}$ through reactions of OH 
    with H$_2$. These appear to be the ideal conditions for the
    presence of large quantities of \hdo\ in the (circum)nuclear regions of
  galaxies.

\begin{acknowledgements}
We are very grateful to Chentao Yang for useful discussions on the data
reported in Y13. 
E.G-A is a Research Associate at the Harvard-Smithsonian
Center for Astrophysics, and thanks the Spanish 
Ministerio de Econom\'{\i}a y Competitividad for support under projects
  AYA2010-21697-C05-0 and FIS2012-39162-C06-01. 
Basic research in IR astronomy at NRL is funded by 
the US ONR; J.F. also acknowledge support from the NHSC.
This research has made use of NASA's Astrophysics Data System (ADS) 
and of GILDAS software (http://www.iram.fr/IRAMFR/GILDAS).
\end{acknowledgements}

\begin{appendix}
\section{Two opposite, extreme cases: Arp 220 and NGC 1068}
\label{appa}

Arp~220 and NGC~1068 are prototypical sources that have been observed
  at essentially all wavelengths. With regard to their \hdo\ submm emission,
  these galaxies are extreme cases and deserve special consideration. 

\begin{figure}
   \centering
   \includegraphics[width=8.5cm]{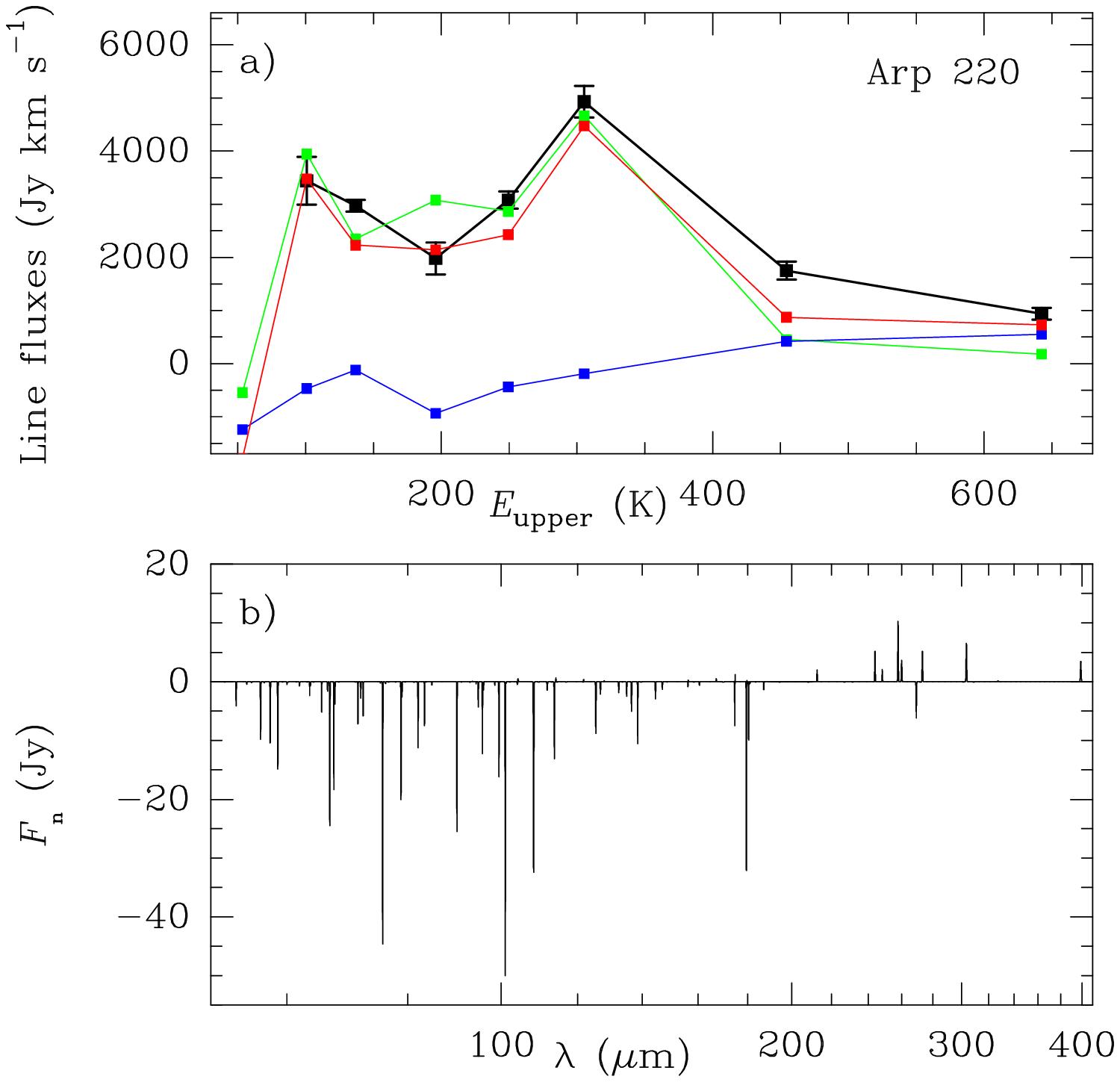}
   \caption{{\bf a)} Proposed composite model for the \hdo\ submm
     lines in Arp~220 (see G-A12), compared with the observed line fluxes
     \citep[black squares, from][]{ran11}. Toward the far-IR optically
     thick nuclear region (blue symbols), the $E_{\mathrm{upper}}<400$ K lines
     are expected mostly in 
     absorption. The \hdo\ emission is generated around that nuclear region,
     in the \cext\ (G-A12) component (green). Red is total. {\bf b)} 
     Resulting predicted 
     composite PACS/SPIRE \hdo\ continuum-subtracted spectrum of Arp~220,
     which is dominated by absorption of the continuum.
   }   
    \label{arp220}
    \end{figure}

In the nearby ULIRG Arp~220, discrepancies between the observed SLED
\citep[][Y13]{ran11} and the single-component models of Fig.~\ref{mod}a1-c1
are worth noting. The observed high $L_6/L_{\mathrm{IR}}\approx2.4\times10^{-5}$
(Fig.~\ref{f25f60}), together with the high 6/2 ratio of $\approx1.4$
(Fig.~\ref{arp220}a), suggest $T_{\mathrm{dust}}\gtrsim65$ K and
$N_{\mathrm{H_2O}}\gtrsim10^{17}$ \cmd, consistent with detection of lines 
$7-8$. However, high \tdust\ and $N_{\mathrm{H_2O}}$ are mostly compatible
with $F_4/F_3>1$, while the observed ratio is $\approx0.7$
(Fig.~\ref{arp220}a). As in Mrk~231, a composite 
model is required to account for the \hdo\ SLED in this galaxy.

In sources with very optically thick and very warm cores such as Arp 220
(G-A12), the increase in $\tau_{100}$ above 1 decreases the submm \hdo\ fluxes
due to the rise of submm extinction (Fig.~\ref{tau100}). While higher \tdust\
generates warmer SEDs, but lowers the $L_{\mathrm{H_2O}}/L_{\mathrm{IR}}$
ratios for lines $2-6$, the increase in $\tau_{100}$ further decreases
$L_{\mathrm{H_2O}}/L_{\mathrm{IR}}$. This behavior suggests that the optimal
environments for efficient \hdo\ submm line emission are regions with high
far-IR radiation density but moderate extinction, i.e., those that {\em
  surround} the thick core(s) where the bulk of the continuum emission is
generated. In contrast, the \hdo\ absorption at shorter wavelengths is more
efficiently produced in the near-side layers of the 
optically thick cores, primarily if high-lying lines are involved. Absorption
and emission lines are thus complementary, providing information on
the source structure. 

We have taken the models in G-A12 for Arp~220 to predict its submm \hdo\
emission. In Fig.~\ref{arp220}a, the blue symbols/line indicate the 
predicted \hdo\ fluxes towards the optically thick, warm nuclear region
(both \cwest\ and \ceast, see G-A12), indicating that most submm lines (with
the exception of lines 3, 7, and 8) are predicted in
absorption. The observed \hdo\ submm line emission \citep{ran11}
must therefore arise in the surrounding, optically thinner region,
i.e., the \cext\ component, where the \hdo\ abundance in
the inner parts ($R\lesssim150$ pc, where $T_{\mathrm{dust}}=70-90$ K) is
increased relative to G-A12 (so \cext\ has
$N_{\mathrm{H_2O}}=1.3\times10^{17}$ \cmd\ in Fig.~\ref{arp220}a). 
According to our model, the relatively low flux in line 4 is due to line
absorption towards the nuclei. The main drawback of the model in
Fig.~\ref{arp220}a is that line 7 is underestimated by a factor 2.
The submm \hdo\ emission in Arp~220 traces a transition region between the
compact optically thick cores and the extended kpc-scale 
disk (G-A12). The overall \hdo\ spectrum is, however, dominated by absorption
of the continuum (Fig.~\ref{arp220}b).

   \begin{figure}
   \centering
   \includegraphics[width=8.5cm]{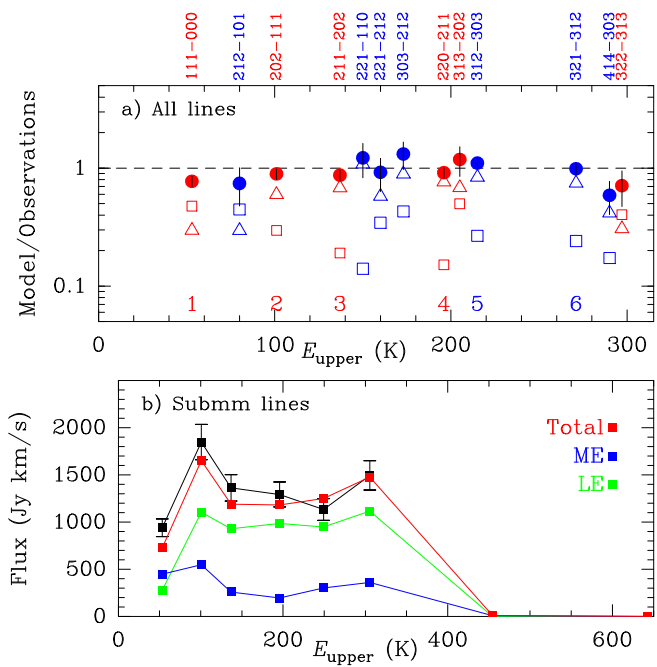}
   \caption{{\bf a)} Composite model for the \hdo\ emission in NGC~1068
     favored in this work. Blue/red indicate ortho/para lines, and the submm
     \hdo\ $1-6$ lines (Table~\ref{tab:tran}) are indicated. Open squares and
     triangles show the contribution by a moderate-excitation (ME) and a
     low-excitation (LE) component, and filled symbols indicate the total
     emission (see text for details); in both cases the submm \hdo\ lines
     $2-6$ are pumped through far-IR photons emitted by dust.
     {\bf b)} Comparison between the observed fluxes of the \hdo\ submm lines
     \citep[black squares, from][]{spi12} and those predicted with the
     composite model.  
   }   
    \label{ngc1068}
    \end{figure}

Just the opposite set of conditions characterizes the nearby
Seyfert 2 galaxy NGC~1068, since the nuclear continuum emission is optically
thin and collisional excitation is important (S12). All detected \hdo\ lines,
including those in the far-IR ($100-200$ $\mu$m) are seen in emission, and
most of them show fluxes (in erg/s/cm$^2$) unrelated to wavelength, upper level
energy (up to $\approx300$ K), or $A$-Einstein coefficient (S12). In
particular, the \hdo\ \t221110\ (108 $\mu$m) and \t221212\ (180 $\mu$m) lines
share the same upper level and show similar fluxes but the $A$-Einstein
coefficient of the 108 $\mu$m transition is a factor of $8.4$ higher than that 
of the 180 $\mu$m transition. With pure collisional
excitation, the only way to account for the observed line ratios is to
invoke high densities and \hdo\ column densities, but also a relatively low
\tgas\ to avoid significantly populating the high-lying levels
($>300$ K). S12 found that $T_{\mathrm{gas}}\sim40$ K, and very high
$N_{\mathrm{H_2O}}$ and $n_{\mathrm{H_2}}$ can provide a reasonable fit to the
SLED. However, these conditions are unrelated to
the warmer gas conditions in the nuclear region of
NGC~1068, as derived from the CO SLED \citep[S12,][hereafter H12]{hay12}.
In addition, the observed \hdo\ submm SLED (Fig.~\ref{ngc1068}) is fairly
similar to the SLEDs obtained in optically thin models with significant
collisional excitation of the low-lying levels.

We have explored an alternative composite solution for the
\hdo\ emission in NGC~1068 with lower densities and \hdo\ columns
and higher \tgas, based on the far-IR pumping of the lines by an external
{\it anisotropic} radiation field. In this framework, we can account for the
weakness of the 108 $\mu$m line by the absorption of continuum
photons, and indeed we would have to explain why this line is not observed to
be even weaker than it is or in absorption. The higher lying far-IR \t322313\
emission line at $156.2$ $\mu$m is in this scenario pumped through absorption
of continuum photons in the \t322211\ line at 90 $\mu$m. 

For the first component, we closely follow H12 in modeling the  
moderate-excitation (ME) component as an
ensemble of clumps, which are described by
$T_{\mathrm{dust}}=55$ K, $\tau_{100}=0.18$, $n_{\mathrm{H_2}}=10^6$ \cmt,
$T_{\mathrm{gas}}=150$ K, and $N_{\mathrm{H_2O}}=6.5\times10^{16}$ \cmd, and
$V_{\mathrm{turb}}=15$ \kms\ (giving $K_{\mathrm{vir}}\sim10$, see H12).
With a mass of $7.5\times10^6$ \Msun, this component is unable to account for
the \hdo\ submm lines $2-6$, but generates a significant fraction of the
observed emission in line 1 and some far-IR lines (Fig.~\ref{ngc1068}a and panel b).

We then added another, low-excitation (LE) component, which is
  identified with the gas generating the low-$J$ CO lines
  \citep[][S12]{kri11} and is thus assigned a density of 
$n_{\mathrm{H_2}}=2\times10^4$ \cmt. For simplicity, we also assume 
$T_{\mathrm{dust}}=55$ K, $\tau_{100}=0.18$, and
$N_{\mathrm{H_2O}}=6.5\times10^{16}$ \cmd\ as for the ME, but adopt the higher
$V_{\mathrm{turb}}$ of $60$ \kms\ (giving $K_{\mathrm{vir}}\sim7$).
For the LE component, and besides the {\em internal} far-IR field described by
its $T_{\mathrm{dust}}$ and $\tau_{100}$, we also follow H12 in
including an {\em external} field (associated with the emission from the whole
region), which is described as a graybody with $T_{\mathrm{BG}}=55$ K and
$\tau_{100}^{BG}=0.05$. The resulting mean specific intensity at 100 $\mu$m of
the external field, $J_{\mathrm{ext}}^{100\mu\mathrm{m}}$, matches the value estimated by H12
within a factor of 2 (their Eq.~1). 
A crucial aspect of the present approach is that this external field is
assumed to be anisotropic, that is, it does not impinge into the LE clumps on
the back side (in the direction of the observer). As a result, the external
field contributes to the \hdo\ excitation without generating absorption in the
pumping far-IR lines (though some absorption is nevertheless produced by the
internal field). As shown in Fig.~\ref{ngc1068}a, the LE  
component is expected to dominate the emission of the submm lines $2-6,$ as
well as the emission of the majority of the far-IR lines. The required mass of
the LE component is $3.5\times10^7$ \Msun, consistent with the mass
inferred from the CO lines for the CND (S12), and the IR luminosity is
$2.6\times10^{10}$ \Lsun.

A key assumption of the present model is that the external radiation field
  does not produce absorption in the far-IR lines, as otherwise (that is, in a
  perfectly isotropic radiation field) the strengths of the far-IR lines would
  weaken, and in particular, the \hdo\ \t221110\ line at 108 $\mu$m line would
  be predicted to be observed in absorption. The proposed anisotropy could be
  associated with the heating by the central AGN, and it seems
  possible as long as the source is optically thin in the far-IR. Radiative
  transfer in 3D would be required to check this feature. On the other hand,
  the external field, while having an important effect on the far-IR
  lines, has a secondary effect on the submm lines, which are
  primarily pumped by the internal (isotropic) radiation field (that is, by
  the dust that is mixed with \hdo). With the caveat of the assumed intrinsic
  radiation anisotropy in mind, we preliminary
  favor this model over the pure collisional one in predicting the \hdo\ submm
  fluxes and conclude that radiative pumping most likely plays an important
  role in exciting the \hdo\ in the CND of NGC~1068.

From the models for these two very different sources and the case of Mrk~231
studied previously (G-A10), we conclude that the excitation of the submm 
\hdo\ lines other than the \t111000\ one is dominated by radiative pumping,
though the relatively low-lying \t202111\ line may still have a significant
``collisional'' contribution in some very warm/dense nuclear regions,
and the radiative pumping may be enhanced with collisional excitation 
of the low-lying $1_{11}$ and $2_{12}$ levels. These individual cases also show
that composite models to account for the full \hdo\ far-IR/submm spectrum
in a given source may be a rather general requirement.

\end{appendix}

\end{document}